\documentclass[aps,prl,superscriptaddress,twocolumn]{revtex4-2}
\usepackage{comment}
\usepackage[utf8]{inputenc}
\usepackage{amssymb}
\usepackage{amsmath}
\usepackage{amsfonts}
\usepackage{bm}
\usepackage{mathtools}
\usepackage{bbold}
\usepackage{upgreek}
\usepackage{xfrac}
\usepackage{soul}
\usepackage{bm}
\usepackage{xcolor}
\usepackage{url}
\usepackage{txfonts}
\usepackage{bbm}
\usepackage{dsfont}
\usepackage{physics}
\usepackage{svg}
\usepackage[nolist]{acronym}

\DeclareSymbolFont{matha}{OML}{txmi}{m}{it}
\DeclareMathSymbol{\varv}{\mathord}{matha}{118}

\begin{document}

\title{Vortex Detection from Quantum Data}

\author{Chelsea A. Williams}
\affiliation{Department of Physics and Astronomy, University of Exeter, Stocker Road, Exeter EX4 4QL, United Kingdom}
\affiliation{PASQAL, 7 Rue Léonard de Vinci, 91300 Massy, France}
\author{Annie E. Paine}
\affiliation{PASQAL, 7 Rue Léonard de Vinci, 91300 Massy, France}
\author{Antonio A. Gentile}
\affiliation{PASQAL, 7 Rue Léonard de Vinci, 91300 Massy, France}
\author{Daniel Berger}
\affiliation{Siemens AG, Gleiwitzer Str. 555, 90475 N\"urnberg, Germany}
\author{Oleksandr Kyriienko}
\affiliation{School of Mathematical and Physical Sciences, University of Sheffield, Sheffield S10 2TN, United Kingdom}
\date{\today}

\begin{abstract}
Quantum solutions to differential equations represent quantum data --- states that contain relevant information about the system's behavior, yet are difficult to analyze. We propose a toolbox for reading out information from such data, where customized quantum circuits enable efficient extraction of flow properties. We concentrate on the process referred to as quantum vortex detection (QVD), where specialized operators are developed for pooling relevant features related to vorticity. Specifically, we propose approaches based on sliding windows and quantum Fourier analysis that provide a separation between patches of the flow field with vortex-type profiles. First, we show how contour-shaped windows can be applied, trained, and analyzed sequentially, providing a clear signal to flag the location of vortices in the flow. 
Second, we develop a parallel window extraction technique, such that signals from different contour positions are coherently processed to avoid looping over the entire solution mesh. We show that Fourier features can be extracted from the flow field, leading to classification of datasets with vortex-free solutions against those exhibiting Lamb-Oseen vortices. Our work exemplifies a successful case of efficiently extracting value from quantum data and points to the need for developing appropriate quantum data analysis tools that can be trained on them. 
\end{abstract}

\maketitle

\begin{acronym}
  \acro{AUC}{area under the curve}
  \acro{CFD}{computational fluid dynamics}
  \acro{CNN}{convolutional neural network}
  \acro{HEA}{hardware efficient ansatz}
  \acro{ML}{machine learning}
  \acro{MSE}{mean squared error}
  \acro{PDE}{partial differential equation}
  \acro{QCNN}{quantum convolutional neural network}
  \acro{QFT}{quantum Fourier transform}
  \acro{QML}{quantum machine learning}
  \acro{QDNN}{quantum deep neural network}
  \acro{QVD}{quantum vortex detection}
  \acro{SNR}{signal-to-noise ratio}
  \acro{TVD}{total variation distance}
\end{acronym}

\section{I. Introduction}

Turbulent flows are inherently nonlinear and exhibit a rich hierarchy of spatiotemporal structures across scales, ranging from large-scale eddies down to dissipative microvortices \cite{Kolmogorov1941,Robinson1991,Jimenez1993,OConnor2016}. Vortex detection is a critical process in computational fluid dynamics (CFD) \cite{anderson1995computational} and is essential for understanding the physics of flow fields \cite{Hussain1986,Jeong1995,coherent_structures}. Accurate detection enables better turbulence modeling and multi-scale analysis, leading to improved predictions in the areas of magnetohydrodynamics \cite{Shibata2011}, astrophysics \cite{Goldreich1995,MacLow2004}, oceanography \cite{Couston2020}, aerospace \cite{McWilliams1984,Spalart1994}, and atmospheric science \cite{Smagorinsky1963}. For industrial and engineering applications, vorticity structures provide key insights to optimize designs with respect to energy efficiency, noise, and safety \cite{cfd_turbine_design,cfd_noise_study,Vermeer2003,Lighthill1952,Bearman1984}. However, it is often challenging to resolve such vortical structures unless fine-mesh direct numerical simulation (DNS) is performed \cite{Moin1998,Kaneda2003,Lee2015}.

Quantum computing methods have the potential to implement DNS-type methods using a distinct physical hardware and large operational space \cite{harrow2009quantum,algorithms_survey}. Quantum algorithms can encode solutions on a fine mesh or function basis into the Hilbert space of a quantum system \cite{Cao_2013,Montanaro2016,Xin2020,MartinSanz2023,Williams2023,Kyriienko2024protocols}. Recent advances in quantum differential equation solvers include methods based on quantum signal processing \cite{Lin2020optimalpolynomial,Linden2022quantumvsclassical,Krovi2023improvedquantum}, eigenstate filtering-based approaches \cite{Lin2020optimalpolynomial,costa2023improving}, linear combinations of unitaries (LCU) \cite{childs2017quantum,Childs2012,Berry2017,jennings2023efficient,Gribling2024}, Schrödingerization~\cite{jin2022quantum,jin2022quantumEXT,jin2023analog,jin2023quantum,hu2024quantum}, Fourier transform-based solvers \cite{Kyriienko2024protocols,LiuCirak2024,lubasch2025fourier,devereux2025drift}, digitized quantum adiabatic methods \cite{subacsi2019quantum,Costa2022,DongLin2022}, and quantum iterative solvers \cite{williams2024iterative,quantum_iterative_solver,time_marching_multigrid,jin2023quantum,quantum_multigrid}. Nonlinearity can be treated with Carleman linearization \cite{JPLiu2021,Krovi2023improvedquantum,costa2023improving,tanaka2023carleman,ingelmann2023quantum,wu2024quantum,gonzalezconde2024,Sanavio2024,sanavio2024carleman}, Chebyshev-based models \cite{Paine2023,wu2025heff}, and quantum nonlinear processing units \cite{lubasch2020variational,jaksch2022variational, pool2024nonlinear}. Turbulence from the quantum encoding perspective was studied in Refs.~\cite{gonzalezconde2024}. However, the drawback of representing solutions in an amplitude-encoded way corresponds to a large readout cost --- we cannot readily access information from quantum states and need to rely on sampling (repeated measurements) \cite{Biamonte2017}. This calls for specialized feature extraction methods for quantum data \cite{Williams2024readout}.

Vortex detection from a calculated flow profile can combine local and global approaches \cite{swirl_vortex_identification,Epps2017}, where the latter is computationally intensive and relies on identifying topological features. In the domain of data-driven science and engineering, machine learning (ML) demonstrated success in flow analysis \cite{Brunton2020rev,Duraisamy_annurev} through advances in pattern recognition, classification, and segmentation. Convolutional neural networks (CNNs) and U-Net architectures were adapted to vortex detection tasks by utilizing convolutional filters \cite{cnn_vortex_identification, vortex_seg_net, vortex_u_net, vortex_cnn, vortex_elm_net}. However, these methods act on classical flow field images or simulation data arrays, and are not applicable for quantum states. Quantum machine learning (QML) developed as a field where quantum datasets can be processed with parameterized quantum circuits \cite{Huang2022,schatzki2021entangled,Caro2022}, spotting patterns that are difficult to analyze classically \cite{Huang2022,Anschuetz2023,umeano2024geometric,umeano2024forrelation,lewis2025functions,morohoshi2025hamiltonians,barthe2025dynamics}, as well as developing physics-informed quantum processing methods \cite{kyriienko2021solving,Paine2021,paine2023quantum,Markidis2022qpinns,heim2021quantum,kasture2022protocols,pinn_qml,Jaderberg2024,Jaderberg2024weather,delejarza2025QCPM,wu2025heff}. From the feature selection perspective, quantum convolutional neural networks became a tool for efficient readout \cite{Cong2019,pesah2021absence,Hur2022,SChen2022,herrmann2022realizing,Zapletal2024,Umeano2024QCNN,Gil-Fuster2024,Song2024frontiers}. At the same time, their utility largely depends on data, competition with measure-first approaches \cite{goh2023liealgebraic,Cerezo2023CSIM,bermejo2024qcnn}, and generally suffers from being physics-agnostic. Our recent work introduced quantum scientific machine learning (QuaSciML) tools for feature extraction from quantum PDE solvers \cite{Williams2024readout}, highlighting the need for specialized post-processing pipelines that operate within the quantum domain. These tools must not only preserve the computational advantages of quantum encoding but also support the efficient retrieval of physically meaningful observables.
\begin{figure*}
    \centering
    \includegraphics[width=1.0\linewidth]{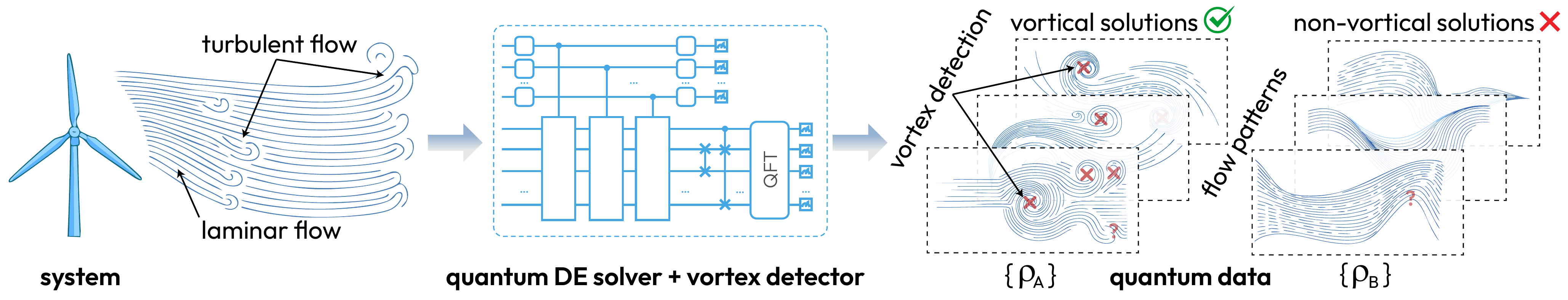}
    \caption{\textbf{Workflow for vortex detection from quantum data.} A nonlinear system is modeled using a quantum differential equation solver, producing states that contain solutions with vortices (turbulent flow) or without vortices (laminar flow). Vortex detection is applied with sliding window-based spectral analysis, compressing relevant features of the flow for consequent detection. The detection circuit is trained on few examples and only requires several tuning parameters.}
    \label{fig:workflow}
\end{figure*}

In this work, we address the challenge of extracting flow features, and specifically vortex patterns, from quantum data produced by quantum differential equation solvers. We propose a quantum vortex detection framework based on sliding window operators paired with quantum Fourier analysis. Our method enables spatially localized probes of the vorticity field and parallelized processing of solution patches, avoiding full quantum state tomography. In this work we show that a general strategy for extracting structured information from quantum PDE solutions requires physics-aware approaches and models with only few tunable parameters.

\section{II. Algorithms} \label{Algorithms}

We proceed to propose and develop tools for vortex detection from quantum data, referred to as \acf{QVD}. The methodology employed for the \ac{QVD} framework takes inspiration from the ideas behind convolutional neural networks. CNNs rely on pattern recognition using a sliding window that traverses a data sample (array of values) to identify local features and hierarchical patterns within the data \cite{oshea2015cnns}. Similarly, we design a \ac{QVD} framework for quantum feature extraction to detect vortical structures in local patches of field profiles that are analyzed in superposition. The designed models are able to learn on patterns inferred from available simulations or data. By working explicitly with quantum simulations (data), the \ac{QVD} model bypasses classical-to-quantum conversion for post-processing. Unlike general quantum convolutional neural networks (QCNN) \cite{Cong2019,pesah2021absence,Umeano2024QCNN,bermejo2024qcnn} that simply mimic the convolution-pooling structure, the \ac{QVD} framework is designed to analyze the Fourier representation of quantum data, as well as the underlying power spectrum. It is a specialized model that embeds principles from \ac{CFD} into the automated detection of vortices. Such a physics-informed methodology and superposition-based processing cannot be replicated within a traditional CNN architecture.

Let us proceed to describe the input for quantum vortex detection. This consists of quantum states as solutions of quantum differential equations. At the physical level they contain local multi-dimensional patches of vorticity field solutions that are amplitude-encoded by construction. As a particular example we consider the Lamb-Oseen vortex model \cite{swirl_vortex_identification}, generating solutions of the type shown in Eq.~\eqref{eq:lamb} for varying parameters. As a solution to the Navier-Stokes equations in cylindrical coordinates, Lamb-Oseen exhibits vortical motion with a concentrated vorticity that diffuses over time \cite{lamb_oseen_vortex}. The model characterizes a two-dimensional vortex structure with a Gaussian distribution of vorticity. The azimuthal velocity profile $v_\theta$ is given by 
\begin{equation}
    v_\theta = v_{\rm{max}} \biggl( 1+\frac{1}{2\Delta} \biggr) \frac{r}{R} \biggl[ 1 - \text{exp} \biggl(-\Delta\frac{R^2}{r^2}\biggr) \biggr],
    \label{eq:lamb}
\end{equation}
where $\Delta$ is the vorticity diffusion parameter, $v_{\rm{max}}$ is the peak tangential velocity, $R$ is the radial distance from the vortex center and $r$ is the radius of the vortex core where the vorticity is concentrated. We take this as an illustrative example and describe the approach to study solutions.
%



\subsection{Quantum vortex detection: tools}

\begin{figure}[t!]
    \centering
    \includegraphics[width=0.8\linewidth]{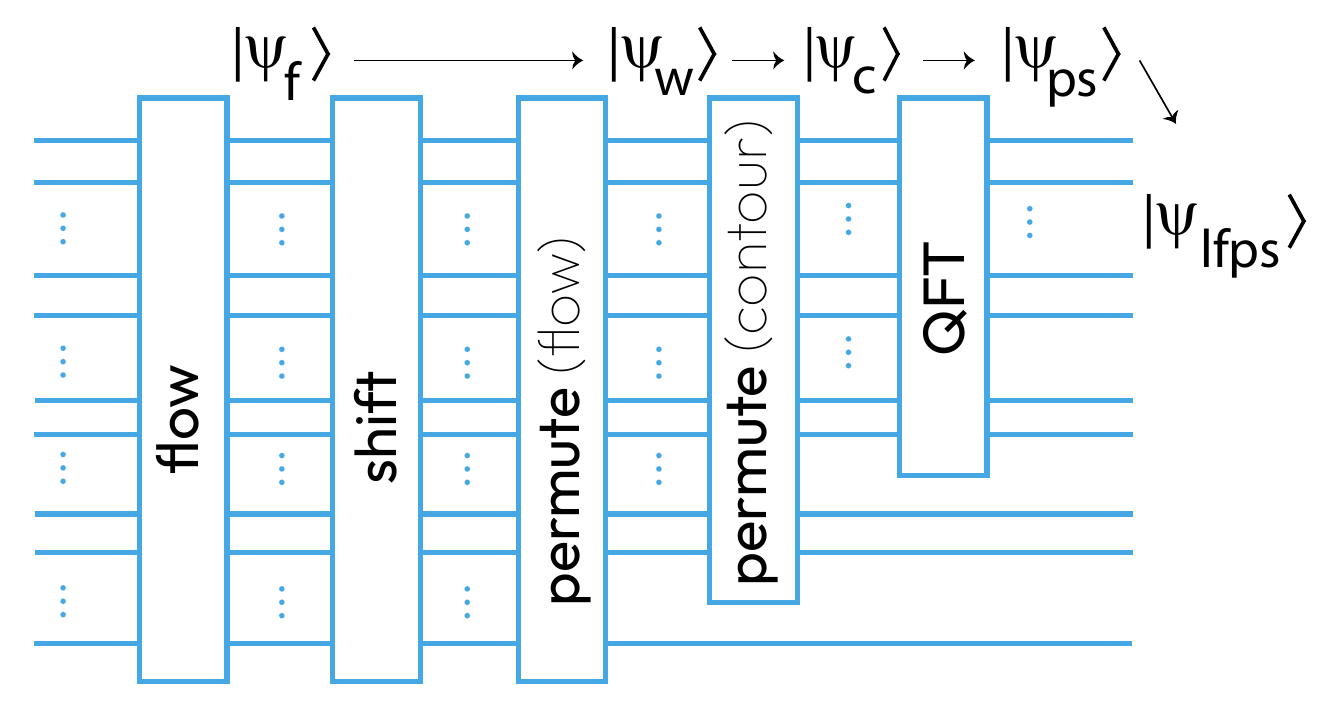}
    \caption{\textbf{Quantum circuit used to perform feature extraction in \ac{QVD}.} Quantum data in the form of a multi-dimensional flow field is encoded into state $\ket{\psi_{\rm{f}}}$. A shift gate followed by a permutation gate are applied to select a window from this flow whose elements are encoded into the top register of state $\ket{\psi_{\rm{w}}}$. Another permutation gate is applied to select a circular contour from the window whose ordered elements are encoded into the top register of state $\ket{\psi_{\rm{c}}}$. The QFT is applied to calculate the power spectrum of the contour, resulting in state $|\psi_{\rm{ps}}\rangle$. The low-frequency band of this power spectrum is extracted from state $|\psi_{\rm{lfps}}\rangle$ by measuring the bottom register of the circuit. This circuit assumes that $n_{\rm{f}} \geq n_{\rm{w}} > n_{\rm{c}} > n_{\rm{lfps}}$. 
    }
    \label{fig:generic_quantum_circuit}
\end{figure}

The input corresponds to the flow field as a quantum state, $\ket{\psi_{\rm{f}}}$. We consider this a starting point for QVD. This is followed by a sequence of operations that process states (Fig.~\ref{fig:generic_quantum_circuit}), which we introduce below. The goal is to perform either a regression or classification task, e.g. assigning a degree of vorticity or classifying turbulent vs laminar flow. Given that we sequentially study patches, we can typically infer the number of vortices, alongside their locations, at the expense of repeated circuit evaluations.

The quantum vortex detection circuit requires the following operations represented by individual blocks in Fig.~\ref{fig:generic_quantum_circuit}. Starting from a $n$-qubit input state, a shift block corresponds to
\begin{equation}
    S(n,d) = \sum_{j=0}^{2^n -1} |(j+d) \text{ mod } 2^n\rangle\langle j|,
    \label{eq:shift}
\end{equation}
which cyclically shifts the computational basis states by a fixed offset $d$. Mathematically, operations in Eq.~\eqref{eq:shift} correspond to generalized Pauli matrices \cite{ramakrishnan1971generalized}. These are closely related to Toeplitz matrices that can be block-encoded efficiently \cite{Sunderhauf2024blockencoding}. The shift operator was shown to scale logarithmically with sparsity and requires $O(\mathrm{poly}(n))$ two-qubit gates coming from the decomposition of multi-controlled (Toffoli-like) gates \cite{camps2023explicitquantumcircuitsblock}. These operators were shown to compile efficiently (linearly in system size) with parallel implementation strategies \cite{Budinski_2023}. Note that powers of shift operators also can be compiled efficiently with $O(n^2)$ complexity \cite{camps2023explicitquantumcircuitsblock}.

The second step of the QVD circuit corresponds to permutation blocks that structure our quantum data into a suitable form. A permutation operation can be written as 
\begin{equation}
    P(n,l) = \sum_{j=0}^{2^n -1} |j\rangle\langle \text{o}_l[j]|,
    \label{eq:permute}
\end{equation}
which permutes the set of computational basis states $\mathcal{H} = \{|j\rangle\}_{j=0}^{2^n-1}$ according to the array $\text{o}_l$, such that order $\text{o}_l[j]$ specifies the index of the state mapped to $|j\rangle$. This represents a bijective map and correspondingly a unitary operator. Exact compilation of $P(n,l)$ depends on the system and the required order. While the general complexity of permuting bitstrings is exponential in the worst case \cite{Shende2006}, we only care about permuting a subset of states $\mathcal{S} = \{ |j\rangle^{(s)}\}$, and do not care about the $\mathcal{H} - \mathcal{S}$ subset. In this case, permutations scale with the cardinality of the relevant subset $d_\mathcal{S}=|\mathcal{S}|$, requiring $O(d_\mathcal{S} n)$ operations \cite{Herbert2024}. As our goal is to bring the computational space into a block-type structure, the complexity can also be inferred from the quantum Schur transform which is polylogarithmic in system size \cite{wills2024schur}.

Finally, the \ac{QFT} is added to enable processing over the low-frequency spectrum rather than a wide-support quasi-probability distribution. The corresponding bijective operator reads as 
\begin{equation}
    \text{QFT}(n) = \Bigl(1/\sqrt{2^n}\Bigr) \sum_{j,k=0}^{2^n -1} \exp\left[-2\pi i (jk/2^n)\right]|k\rangle\langle j|,
    \label{eq:QFT}
\end{equation}
with $\Omega= \{ \ket{k} \}_{k=0}^{2^n-1}$ denoting the frequency basis states. The corresponding map $\text{QFT}: \Omega \mapsto \mathcal{H}$ and its reverse can be implemented with $O(n^2)$ two-qubit gates in its textbook form \cite{nielsen2010quantum} and compiled efficiently to various gatesets. Approximate versions of QFT can also be run in linear gate complexity and logarithmic depth \cite{baumer2025aqft}. 

One more ingredient is the $k$-frequency projector $\Pi(m) = \sum_{k=0}^{2^m -1} |k\rangle\langle k|$. This is a rank-$2^m$ projector from the full $2^n$-dimensional Hilbert space. It can be implemented by measuring the first $n-m$ qubits and post-selecting on $\ket{0}^{\otimes (n-m)}$ \cite{measurement_with_postselection}, or using a coherent ancilla-based projection.


\subsection{Sequential QVD: example}

\begin{figure}[t]
    \centering
    \includegraphics[width=\linewidth]{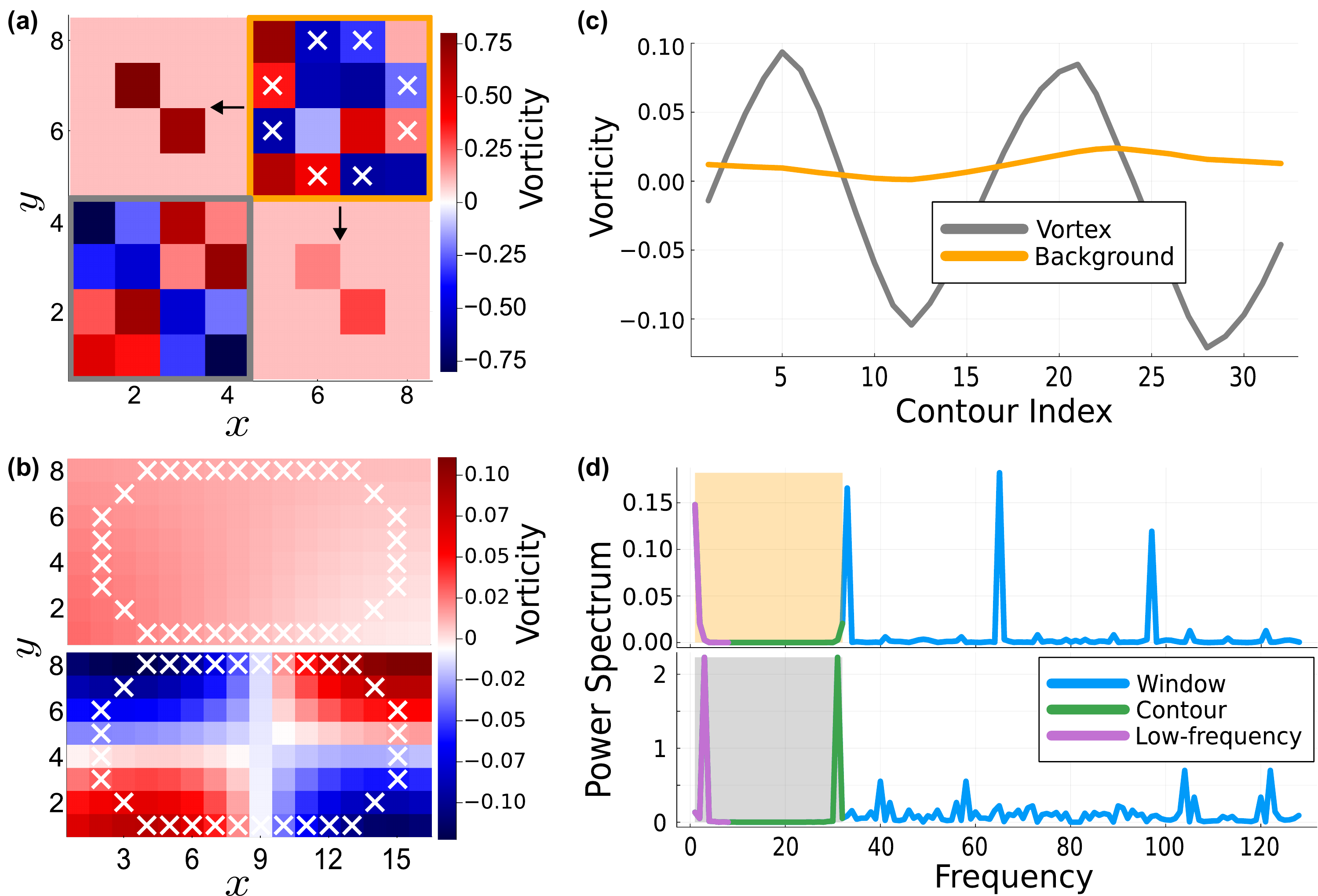}
    \caption{\textbf{Examples of performing feature extraction in \ac{QVD}.} \textbf{(a)} A two-dimensional vorticity flow in the $xy$-plane. A sliding window is selected and the contour points within each window (indicated by white crosses) are extracted. Note that this example is course-grained for brevity. \textbf{(b)} Two $7$-qubit windows are shown with $5$-qubit circular contours extracted. The top window is representative of a featureless background and the bottom window is representative of a centrally-located vortex. \textbf{(c)} The vorticity amplitudes extracted from the two contours. \textbf{(d)} The power spectra of the background and vortex, where frequencies are indexed by window pixel $\psi_{i,j}$. The power spectra is shown for the encoded window ($\rm{QFT}\ket{\psi_{\rm{w}}}$ on $7$ qubits), contour ($|\psi_{\rm{ps}}\rangle$ on $5$ qubits) and low-frequency band ($|\psi_{\rm{lfps}}\rangle$ on $3$ qubits), as obtained with the quantum circuit from Fig.~\ref{fig:generic_quantum_circuit}. Note that vorticity and power spectrum are measured in arbitrary units.}
    \label{fig:power_spectra}
\end{figure}

We proceed to consider a step-by-step example for the QVD approach, and visualize its element. We start with a two-dimensional vorticity flow field $\psi_{i,j}$ in the $xy$-plane that is encoded into a $n_{\rm{f}}$-qubit quantum state $\ket{\psi_{\rm{f}}}$. The encoding of this flow onto two registers is represented by 
\begin{equation}
    \ket{\psi_{\rm{f}}} = \ket{x,y} = \sum_{i,j=0}^{2^{n_{\rm{f}}}-1} \psi_{i,j} \ket{i}\otimes\ket{j},
\end{equation}
where the basis state $\ket{i}$ stores the column index corresponding to the $x$\nobreakdash-register and basis state $\ket{j}$ stores the row index corresponding to the $y$\nobreakdash-register. This is repeated for each flow to create a quantum dataset $\{ \vb*{\rho} \} = \{|\psi_{{\rm{f}},q}\rangle \langle \psi_{{\rm{f}},q}|\}_{q=1}^N$ consisting of $N$ encoded flow field solutions. The workflow outlining how features are extracted from these flows is shown in Fig.~\ref{fig:power_spectra}. In particular, the workflow is designed to identify rotational symmetries in the states, which signal the presence of vortices.

\par Next, in a similar vein to the \ac{CNN}, a window is selected from the flow domain. A schematic of a sliding window traversing a vorticity flow field is shown in Fig.~\ref{fig:power_spectra}(a). To achieve this, a shift gate $S(n,d)$ is applied to $\ket{\psi_{\rm{f}}}$ to select the desired window. This shift gate is followed by a permutation gate $P(n,l)$ which moves the elements of the window to the top register of the circuit. This results in a state $\ket{\psi_{\rm{w}}}$ which encodes the window in the top $n_{\rm{w}}$ qubits. In the \ac{QVD} model, the step size between consecutive window evaluations is treated as a tunable parameter $\alpha$. This helps reduce the number of circuit evaluations while ensuring that there is sufficient space between windows to detect all vortices. 

Next, a circular contour is selected from each window. An example of a circular contour as extracted from two distinguishable windows is demonstrated in Fig.~\ref{fig:power_spectra}(b); one representing the background and the other representing a vortex. Extracting the contour enables tracing the variation in the magnitude of vorticity while accounting for rotational symmetry, effectively producing a state that encodes the vorticity amplitudes. Analyzing the contour of a vorticity window allows to isolate regions where amplitudes exhibit certain structure. Fig.~\ref{fig:power_spectra}(c) shows a comparison of the vorticity amplitudes of the contours extracted from both the background window and the vortex window. To achieve this, a permutation gate is applied to $\ket{\psi_{\rm{w}}}$ to move the elements of the contour (in either a clockwise or anti-clockwise order) to the top register of the circuit. This results in a state $\ket{\psi_{\rm{c}}}$ which encodes the contour in the top $n_{\rm{c}}$ qubits. When comparing the vorticity amplitudes, the contour from the background window has a fairly flat profile but the contour from the vortex window has a distinguishable oscillatory profile. Since vortices are typically characterized by oscillatory vorticity profiles due to rotational symmetry, examining circular contours provides a focused representation of how vorticity varies around a central point. In the \ac{QVD} model, the inverse contour radius is treated as a tunable parameter $\beta$ to account for variations in vortex size and ensure optimal detection.

Given that features are hidden in oscillatoric patterns, we use QFT for the contour (state $|\psi_{\rm{ps}}\rangle$) and sample the corresponding power spectrum to identify signatures of vortical structures (prominent low-frequency components along the contour). Therefore, an adjustable circuit used to extract features is represented by
\begin{align}
\notag
|\psi_{\rm{ps}}\rangle &= \hat{U}(\alpha,\beta ; n_{\rm{f}},n_{\rm{w}},n_{\rm{c}}) \ket{\psi_{\rm{f}}} \\
 &= \text{QFT}(n_{\rm{c}}) P[n_{\rm{w}},l_2(\beta)] P(n_{\rm{f}},l_1) S[n_{\rm{f}},d(\alpha)] \ket{\psi_{\rm{f}}}.
 \label{eq:psi_ps}
\end{align}
Here, the shift operator selects the position of each window according to the step size $\alpha$, while the permutation operator reorders the data specified by the contour with the inverse radius $\beta$.

Finally, the low-frequency region of the power spectrum is extracted. This can be achieved by performing projective measurements on the bottom part of the register (effectively filtering out high frequencies). This results in a state $|\psi_{\rm{lfps}}\rangle$ which encodes the low-frequency power spectrum in the top $n_{\rm{lfps}}$ qubits, $\mathcal{P} = \big|\Pi(n_{\rm{lfps}}) |\psi_{\rm{ps}}\rangle\big|^2$. More specifically, for contour index $p$ and flow field index $q\in [1,N]$, this power spectrum is given by
\begin{equation}
    \mathcal{P}_{p,q} = \big|\Pi(n_{\rm{lfps}}) \hat{U}_p(\alpha,\beta) |\psi_{{\rm{f}},q}\rangle\big|^2.
\end{equation}
If each flow in the dataset $\vb*{\rho}$ is indexed by $q$, the power spectrum must be computed for every extracted contour $p$ as it moves through the field. Fig.~\ref{fig:power_spectra}(d) shows the power spectra of the different states extracted from the window, with a particular focus on contrasting the low-frequency signals between the background window and the vortex window. Vortex signatures are typically present in the frequency region $[0,N_{\mathrm{lf}}]$, where $N_{\mathrm{lf}}$ is $32$ for our example in Fig.~\ref{fig:power_spectra}(d). 
The top power spectrum of the background has a weak signal with no detectable characteristics in the low-frequency band. In comparison, the bottom power spectrum of the vortex has a sharp peak in the low-frequency band with a maximum value of approximately $2.4$. 

The features in $|\psi_{\rm{lfps}}\rangle$ are analyzed to distinguish between the presence or absence of a vortex. When a vortex is present, there is a peak in the low-frequency power spectrum as demonstrated from Fig.~\ref{fig:power_spectra}(d). A threshold value for the peak is used to classify whether a vortex is present or absent within the window. We set the threshold as a tunable parameter $\gamma$, adjusted based on available data to separate vortex signal from background fluctuations or noise. 
%
%

We apply classical post-processing to eliminate over-counting and identify unique vortex detections. Each circuit applied to each patch returns a binary value of $\mathcal{M}_{p,q}(\alpha,\beta,\gamma)$. The \ac{QVD} model therefore often detects the same vortex multiple times across neighboring contour evaluations. To count unique vortices only, we average regions where multiple detections occur, such that $\mathcal{M}_q(\alpha,\beta,\gamma)$ provides $0$ or $1$ detection values over the set of contour values. This post-processing can also aid in determining the central locations of the vortices.

Notably, the \ac{QVD} model can be integrated into a machine learning framework by treating the three parameters as tunable variables with which to optimize vortex detections, based on known examples of vortices. This leads to the physics-informed \ac{QML} approach, where inductive bias is derived from the analysis of the low-frequency power spectrum and optimized based on physically motivated variables.


\subsection{Parallel \ac{QVD}}

As an important step to avoid repeated QVD application, the quantum vortex detection approach can be parallelized. This step is needed to accelerate the processing of flow field solutions and enable coherent feature selection. Specifically, we suggest processing solutions with a linear combination of positions (shifts) $\sum_x |x\rangle \langle x| \otimes \hat{U}_x(\alpha,\beta)$, selected such that the window positions effectively cover the grid (as typical for quantum phase estimation protocols \cite{nielsen2010quantum}). 
\begin{figure}[b]
    \centering
    \includegraphics[width=0.8\linewidth]{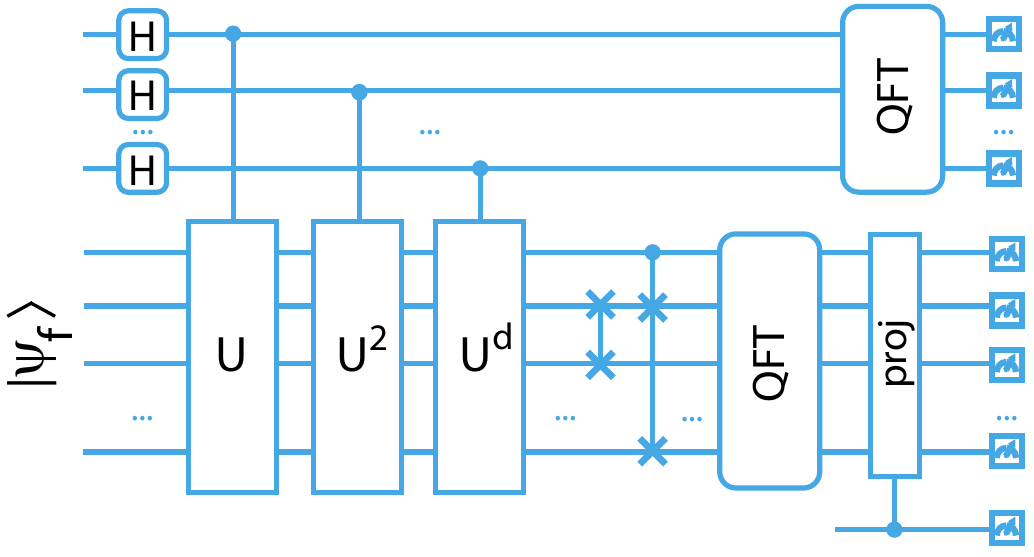}
    \caption{\textbf{Quantum circuit for parallelized feature processing in quantum vortex detection.} Conditioned on the superposition state $|+\rangle^{\otimes n_a}$, operators corresponding to different shifts are applied (each corresponding to associated ancilla state $|x\rangle$), followed by permutations and QFT. Projecting out the bottom register, the low-frequency amplitudes are pushed to the ancilla register, which is to be sampled in the frequency basis.}
    \label{fig:parallel_quantum_circuit}
\end{figure}
Note that the grid can be coarse-grained depending on the typical vortex size. Next, we employ the readout strategy proposed in Ref.~\cite{umeano2024community}, where the ancillary register stores positions in the amplitudes of states $|x\rangle$ and the relevant (low $k$-frequency) component is selected. For instance, this can be achieved by implementing a rank-1 projector $\Pi_k = |k\rangle \langle k|$ and acting on the shifted state as $\Pi_k |\psi_{\rm{ps}}\rangle$. Finally, the QFT is applied on the ancillary register such that we read out information about the spectral (i.e. global) properties of the flow field (Fig.~\ref{fig:parallel_quantum_circuit}). This higher-order spectrum is referred to as the density spectrum. The associated probability distribution $p_{\mathrm{a}}(x)$ can be sampled on the ancillary register, and post-processed to get the global properties of each example.

\begin{figure*}[t]
    \centering
    \includegraphics[width=0.8\linewidth]{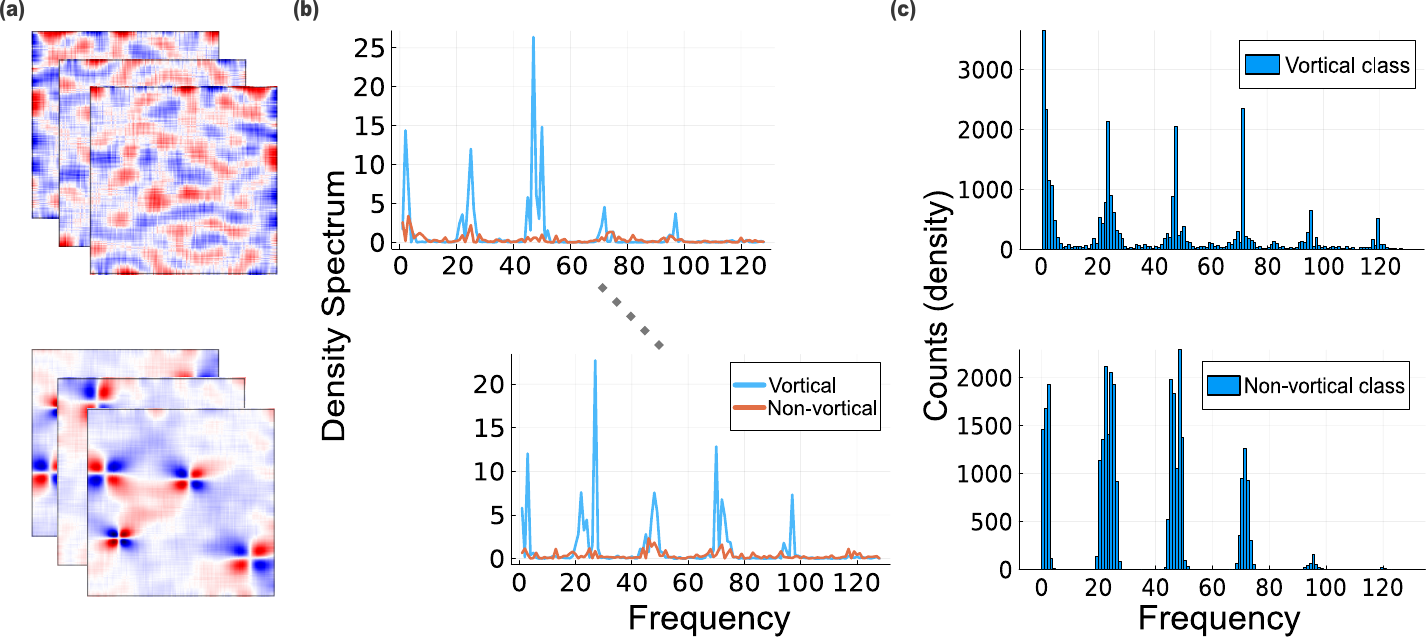}
    \caption{\textbf{Quantum data-driven feature processing in \ac{QVD}.} \textbf{(a)} A quantum dataset containing vorticity flow field solutions $|\psi_{i,j}\rangle$ categorized into non-vortical and vortical classes. \textbf{(b)} The density spectra for each pair of vortical and non-vortical fields, where frequencies are indexed by single power spectrum values of contours $N_{\rm{c}}$ processed in parallel. These spectra are obtained with the quantum circuit from Fig.~\ref{fig:parallel_quantum_circuit} and truncated to $7$ qubits. The top vortical density spectrum corresponds to a field with $4$ vortices and the bottom to one with $7$ vortices. \textbf{(c)} The representative density spectrum distribution of the non-vortical and vortical classes obtained from sampling the individual density spectra of all fields in the dataset. Note that density spectrum is measured in arbitrary units.}
    \label{fig:density_spectra}
\end{figure*}

Let us demonstrate the utility of this approach with the following example. We consider two distinct types of flow fields, separating samples into those with vortical and nonvortical patterns [Fig.~\ref{fig:density_spectra}(a)]. Each quantum data sample is processed coherently, such that selected low-frequency components are pushed to the ancillary register, forming a state that contains information about the entire field (its spectral properties). Next, the QFT is applied on the ancillary register, aiming to highlight states with flat profiles (i.e. frequency response from all positions is similar but noisy) as compared to those with significant $k$-frequency components (i.e. certain areas show pronounced response while other parts of the flow have low signal). The corresponding examples are shown in Fig.~\ref{fig:density_spectra}(b). The density spectrum of a non-vortical field is distinct from that of a vortical field containing any number of vortices. This higher-order spectral analysis is necessary for identifying the density of peaks in the power spectrum $\mathcal{P}(\tilde{k})$, which reflects the underlying vortex density of the flow field solutions.

Given the variability in vorticity, as well as the number, location and strength of vortices, it is useful to compute a representative density spectrum distribution for each flow category. These distributions capture the characteristic spectral features of each class, enabling more robust comparison and feature extraction across different datasets. To construct such distributions, the density spectrum of each field is first computed and then repeatedly sampled (via usual projective measurements). The resulting measurement results are concatenated to form the class-level distributions shown in Fig.~\ref{fig:density_spectra}(c), which are representative of the density spectra produced by fields in each class of the quantum dataset. The non-vortical distribution $p_{\rm{NV}}(x)$ does not exhibit distinctive features, while the vortical distribution $p_{\rm{V}}(x)$ is non-zero only at certain frequencies. 
This shows that the spectral signatures of non-vortical and vortical fields are sufficiently different to enable clear class separation.

\par Access to a representative distribution for each class can thus be leveraged for efficient classification. Empirical distributions can be sampled from their respective representative distributions $p_{\rm{NV}}(x)$ and $p_{\rm{V}}(x)$. 
These empirical distributions can be post-processed by training a classical \ac{ML} classifier that learns the separation between non-vortical and vortical regimes. 
Upon training, we obtain empirical test distributions by sampling the density spectrum of a new field that did not form part of the original dataset composed in Fig.~\ref{fig:density_spectra}(a). This test distribution can then be fed into the classifier, to determine if its underlying field is non-vortical or vortical.

\section{III. Results} \label{Results}

In this section we discuss the use of supervised quantum machine learning, to develop an optimized vortex detection model based on the Lamb-Oseen vortex solution of the viscous Navier-Stokes equations. For simplicity in exhibiting how the \ac{QVD} model is performed, the solutions used for training and testing are obtained from classical simulations using the model from Eq.~\eqref{eq:lamb}. It is a separate challenge to prepare the input data (often with ancilla overheads that prevent testing), and here we concentrate on the readout procedure.  

Vorticity fields for representative Lamb-Oseen solutions are prepared as quantum states and used as inputs to the \ac{QML} model, forming the quantum dataset $\{\vb*{\rho}\}$. It contains vorticity fields, each with grid dimensions of $200\times200$. The two-dimensional velocity fields $\vb*{v}=(v_x,v_y)$ are first generated by randomly placing a set of Lamb-Oseen vortices of different strength, size and orientation onto the grid \cite{swirl_vortex_identification}, achieved by sampling the parameters $\Delta$, $v_{\rm{max}}$ and $r$ in Eq.~\eqref{eq:lamb} as well as randomly selecting the number of vortices $M\in[4,8]$, their location and orientation. We assume that vortices are sufficiently separated to be individually identifiable. We overlay each flow field with a random velocity field defined with a smoothed Gaussian filter to act as noise. The vorticity fields $\psi=\nabla\times\vb*{v}$ are then calculated. 
\begin{figure}[b]
    \centering
    \includegraphics[width=0.65\linewidth]{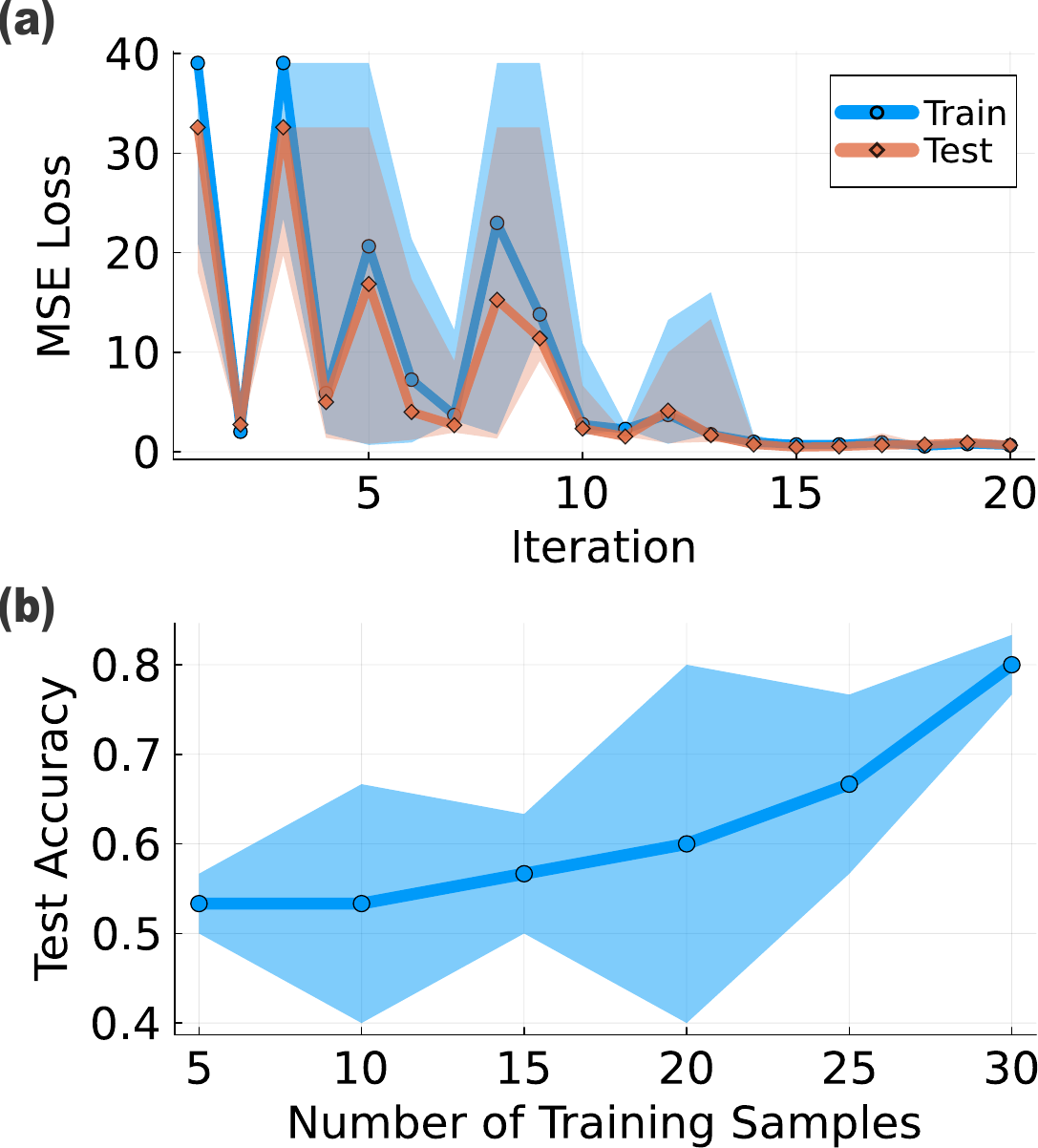}
    \caption{\textbf{Trainability and generalizability of the \ac{QVD} model using Bayesian optimization.} Curves and dots represent the median, and shaded areas represent the inter-quartile range as obtained from $5$ instances with different seed values. \textbf{(a)} Training and testing of the model where we use optimization to minimize the \ac{MSE} with respect to the model's parameters. The loss history is shown for a model trained on $45$ fields and tested on $15$ fields. A final \ac{MSE} loss of $0.6$ for the train set and $0.7$ for the test set is reached after $20$ epochs. \textbf{(b)} The generalizability of the model where optimization is used to maximize the accuracy with respect to the model's parameters. Accuracy is optimized on a growing training set and evaluated on a separate test set of size $30$ containing only unseen fields.}    
    \label{fig:bayesian_optimisation}
\end{figure}

\subsection{Sequential \ac{QVD}}

\par To effectively implement and evaluate the \ac{QVD} model as used for supervised vortex counting, we use an optimization procedure for selected model parameters (window step size, inverse contour radius,  power spectrum threshold) and evaluate performance of trained models. Specifically, we minimize a mean squared error (MSE) loss function that encapsulates the number of correctly detected vortices. The \ac{MSE} is given by
\begin{equation}
    \text{MSE}(\alpha,\beta,\gamma) = \sum_{q=1}^N \frac{[M_q-\mathcal{M}_q(\alpha,\beta,\gamma)]^2}{N},
\end{equation}
where $\mathcal{M}_q$ is the estimated number of unique detections produced by the \ac{QVD} model for the flow field $q$ and parameter set $\{\alpha,\beta,\gamma\}$. Optimization is also performed to maximize the accuracy of detections. The accuracy is defined as the proportion of exact matches between true and predicted vortex counts.
%
%

We show the results obtained using a Bayesian optimization procedure in Fig.~\ref{fig:bayesian_optimisation}. The training stage for the \ac{QVD} model is demonstrated with the \ac{MSE} loss history in Fig.~\ref{fig:bayesian_optimisation}(a). The optimization is summarized over $5$ instances as applied to a sample set of $N=60$ flow fields that is split into a $75\%$ training set and a $25\%$ testing set. The optimization is able to retrieve a \ac{MSE} of less than $1$ in fewer than $15$ epochs. The physically motivated parameters directly influence the sensitivity and accuracy of the vortex detection model, as reflected in the initially high \ac{MSE} that decreases significantly during early optimization stages. The generalization of the \ac{QVD} model is demonstrated with the accuracy in Fig.~\ref{fig:bayesian_optimisation}(b). Here, the optimization procedure identifies the best parameters for vortex detection by maximizing accuracy on an increasing number of training samples. The performance is then measured via the accuracy on the test set which contains a randomized selection of $30$ flow fields not seen during training. This accuracy measures the success rate of vortex detection and how well the optimized parameters of the \ac{QVD} model generalize to unseen data when trained on datasets of increasing size. This result demonstrates how performance differs from large to small training sets and shows that an accuracy of $80\%$ is achievable with as few as 30 fields used for training.
\begin{figure}
    \centering
    \includegraphics[width=0.7\linewidth]{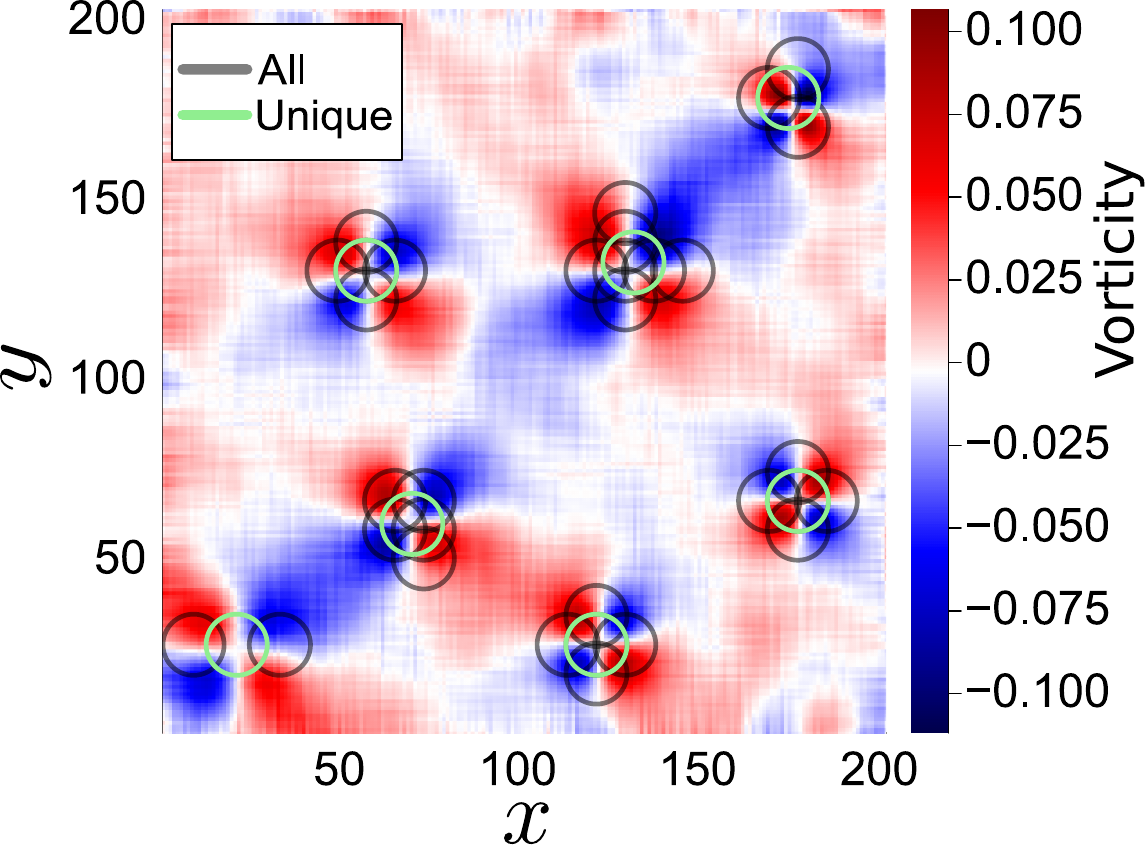}
    \caption{\textbf{Example of the output obtained from the vortex detection model as applied to a Lamb-Oseen flow.} The two-dimensional vorticity flow field is shown in the $xy$-plane and has dimension of $200\times200$. The flow contains $7$ Lamb-Oseen vortices of varying size, strength and orientation positioned randomly on the gird. Black circles represent all detections made by the model, while green circles indicate the final unique detections after averaging across all detections to remove over-counting. The model uses the optimal parameter values consisting of a window step size of $\alpha=8$, a contour with an inverse radius of $\beta=3$ and a power spectrum threshold of $\gamma=0.9$, resulting in a \ac{MSE} of $0.25$.}
    \label{fig:model_detections}
\end{figure}

In Fig.~\ref{fig:model_detections} we show the output of the vortex detection model as applied to an example vorticity field. This vorticity field is derived from a randomly generated Lamb-Oseen velocity field consisting of $7$ vortices. This result is achieved through an exhaustive grid search on all $N$ fields, which identifies the optimal model parameters based on a ground truth \ac{MSE} of $0.25$. The black circles represent the contours where their low-frequency power spectrum peaks above the threshold, as used to indicate the presence of a vortex. The green circles represent the unique vortices, as obtained from averaging over regions where multiple contours overlap. This model output has accurately detected both the number of vortices in the flow field and the central positions of each vortex, demonstrating high accuracy of detection on quantum data. 
\begin{figure}
    \centering
    \includegraphics[width=0.75\linewidth]{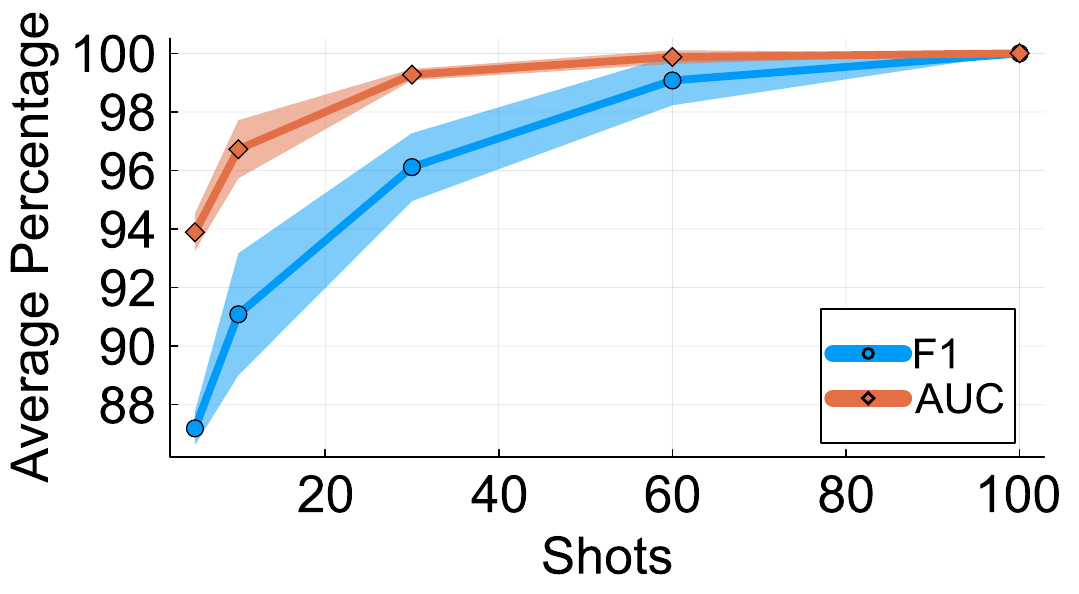}
    \caption{\textbf{Classification from compressed feature representation using spectral density.} Classification is performed on empirical density spectra sampled from the representative distributions of the non-vortical and vortical classes using a random forest classifier. The representative distributions are generated from an equally balanced dataset of $60$ vortical and non-vortical flow fields, with density spectra computed using parallel \ac{QVD} truncated to $8$ qubits. Curves and dots represent the average F1 score and \ac{AUC} metrics and shaded areas represent the standard deviation obtained with $5$-fold cross-validation. The number of samples used to generate the empirical density spectra are given by the number of shots. The number of generated distributions used for classification is given by $10,000$/shots to ensure each model run contains the same number of measurements.}    
    \label{fig:classification}
\end{figure}

\subsection{Parallel \ac{QVD}}

\par To assess the \ac{QVD} algorithm as used for classification, we use classical \ac{ML} models trained on empirical (sampled) density spectra distributions. In this classification task, we aim to separate flow behavior into non-vortical or vortical classes. This can later be used to verify if an unseen density spectrum was generated from a non-vortical or vortical field, enabling predictions of system properties without tomographic procedures. 
Each empirical distribution is sampled with a fixed number of shots. The number of empirical distributions in each set is such that each set contains exactly $10,000$ measurements; when shot size is low, the number of repetitions is high and vice versa. We consider this procedure to provide consistency in assessing classification accuracy as a function of the number of shots. 

In Fig.~\ref{fig:classification} we show results obtained using a standard random forest classifier. The underlying dataset contained $N$ fields, $30$ fields in each class. The non-vortical fields contain no vortices, just random fluctuations, while vortical fields contain $M \in [1,8]$ Lamb-Oseen vortices. Each field is processed using the parallel \ac{QVD} algorithm, with optimal parameters determined by Bayesian optimization as provided in Fig.~\ref{fig:model_detections}. We then compute the density spectrum for each field and truncate it to $8$ qubits. These density spectra are sampled and concatenated to generate representative distributions for each class. We generate the empirical distributions, with sample sets for different shot sizes used to train and test independent classifiers. Each sample set undergoes $5$-fold cross-validation, after which the F1 score and \ac{AUC} are calculated. As expected, the result shows that classification accuracy improves with more measurement shots. Notably, the results indicate that a high F1 score of approximately $90\%$ can be achieved with as few as $5$ shots (for the given system size).

To assess the performance we need to establish baselines. For this, we consider the most typical quantum machine learning approaches based on parametrized quantum circuits (also known as quantum neural networks, QNNs). The \ac{QVD} model demonstrates excellent performance compared to other variational \ac{QML} approaches in the classification of quantum vorticity field data. A comparison in the brute-force classification of quantum data into vortical (turbulent) and non-vortical (laminar) regimes using different methods is shown in Tab.~\ref{tab:comparison}. Each model is trained on a quantum dataset similar to Fig.~\ref{fig:density_spectra}(a), consisting of $N=40$ field solutions, each encoded into quantum states using $16$ qubits. The classification accuracy is evaluated on test sets across $4$ independently generated datasets. In the \ac{QVD} implementation, the representative density spectrum distribution for each class is constructed from all $N$ flows and the subsequent empirical distributions used in classification are sampled with $10$ shots. The benchmark \ac{QML} models include the \ac{QDNN} and the \ac{QCNN} \cite{Cong2019,Umeano2024QCNN}. Each neural network architecture uses a \ac{HEA}, trained using the Adam optimizer with a learning rate of $0.1$ over $50$ iterations. The comparison to a classical \ac{CNN} is not appropriate, since the use of quantum data renders classical approaches inapplicable. The result of this comparison shows the importance of developing bespoke quantum algorithms and the need to design quantum \ac{CFD} workflows that operate natively within quantum simulation environments.
\begin{table}[t!]
\centering
\begin{tabular}{|c|c|}
\hline
\textbf{Method} & \textbf{Test Accuracy (\%)} \\
\hline
QDNN & $63 \pm 30$ \\
QCNN & $63 \pm 32$ \\
QVD & $93 \pm 0.3$ \\
CNN (classical) & not applicable \\
\hline
\end{tabular}
\caption{\textbf{Comparison of different classification methods.} Scores for different approaches (QNN-based and QVD) are summarized for distinguishing turbulent and laminar flows in quantum vorticity field data. Each model is trained on $30$ fields and tested on $10$ fields. Accuracy is measured on the test set, with averages and standard deviations computed over $4$ runs using different seed values. Classical neural networks are not applicable as we work with quantum datasets.}
\label{tab:comparison}
\end{table}


\section{IV. Conclusion}
\label{Conclusion}

We proposed an algorithm for vortex detection on quantum data, where solutions to differential equations are analyzed with circuits of specific structure. The approach, named quantum vortex detection (QVD), relies on extracting features of flow fields via sliding windows (motivated by convolutional approaches), followed by Fourier analysis along contours to identify signals inherent to structures with vorticity. We described the required sequence of operations for shifting, permuting, and transforming states, and demonstrated QVD on examples that encode Lamb-Oseen vortices. In particular, we showed that vortices can be detected with high accuracy when using a few-parameter model trained to set an optimal detection threshold. Using the coherent sample loading via quantum parallelism, we performed classification of vortical vs non-vortical solutions, greatly outperforming other quantum neural network approaches for quantum data.

As direction for future works we consider extensions of these quantum readout tools to other coherent structures, such as eddies, which are often more chaotic and challenging to detect \cite{eddynet}. Also, it would be beneficial to investigate how this detection model can work in a quantum-inspired regime with differential equation solutions stored as a tensor network \cite{Gourianov2022}. Finally, an intriguing possibility for detecting topological flow features is using quantum methods for topological data analysis \cite{Lloyd2016,Gyurik2022,Berry2024,Scali2024dos,Scali2024thermal}, if modified appropriately for the type of data in question.

\section{Acknowledgment}
This work was supported by Siemens Industry Software NV. O.K. acknowledges support from the QCi3 Hub (grant number EP/Z53318X/1).


\begin{thebibliography}{129}%
\makeatletter
\providecommand \@ifxundefined [1]{%
 \@ifx{#1\undefined}
}%
\providecommand \@ifnum [1]{%
 \ifnum #1\expandafter \@firstoftwo
 \else \expandafter \@secondoftwo
 \fi
}%
\providecommand \@ifx [1]{%
 \ifx #1\expandafter \@firstoftwo
 \else \expandafter \@secondoftwo
 \fi
}%
\providecommand \natexlab [1]{#1}%
\providecommand \enquote  [1]{``#1''}%
\providecommand \bibnamefont  [1]{#1}%
\providecommand \bibfnamefont [1]{#1}%
\providecommand \citenamefont [1]{#1}%
\providecommand \href@noop [0]{\@secondoftwo}%
\providecommand \href [0]{\begingroup \@sanitize@url \@href}%
\providecommand \@href[1]{\@@startlink{#1}\@@href}%
\providecommand \@@href[1]{\endgroup#1\@@endlink}%
\providecommand \@sanitize@url [0]{\catcode `\\12\catcode `\$12\catcode `\&12\catcode `\#12\catcode `\^12\catcode `\_12\catcode `\%12\relax}%
\providecommand \@@startlink[1]{}%
\providecommand \@@endlink[0]{}%
\providecommand \url  [0]{\begingroup\@sanitize@url \@url }%
\providecommand \@url [1]{\endgroup\@href {#1}{\urlprefix }}%
\providecommand \urlprefix  [0]{URL }%
\providecommand \Eprint [0]{\href }%
\providecommand \doibase [0]{https://doi.org/}%
\providecommand \selectlanguage [0]{\@gobble}%
\providecommand \bibinfo  [0]{\@secondoftwo}%
\providecommand \bibfield  [0]{\@secondoftwo}%
\providecommand \translation [1]{[#1]}%
\providecommand \BibitemOpen [0]{}%
\providecommand \bibitemStop [0]{}%
\providecommand \bibitemNoStop [0]{.\EOS\space}%
\providecommand \EOS [0]{\spacefactor3000\relax}%
\providecommand \BibitemShut  [1]{\csname bibitem#1\endcsname}%
\let\auto@bib@innerbib\@empty
\bibitem [{\citenamefont {Kolmogorov}(1991)}]{Kolmogorov1941}%
  \BibitemOpen
  \bibfield  {author} {\bibinfo {author} {\bibfnamefont {A.~N.}\ \bibnamefont {Kolmogorov}},\ }\href@noop {} {\bibfield  {journal} {\bibinfo  {journal} {Proc. R. Soc. Lond. A}\ }\textbf {\bibinfo {volume} {434}},\ \bibinfo {pages} {9} (\bibinfo {year} {1991})}\BibitemShut {NoStop}%
\bibitem [{\citenamefont {Robinson}(1991)}]{Robinson1991}%
  \BibitemOpen
  \bibfield  {author} {\bibinfo {author} {\bibfnamefont {S.~K.}\ \bibnamefont {Robinson}},\ }\href@noop {} {\bibfield  {journal} {\bibinfo  {journal} {Annual Review of Fluid Mechanics}\ }\textbf {\bibinfo {volume} {23}},\ \bibinfo {pages} {601} (\bibinfo {year} {1991})}\BibitemShut {NoStop}%
\bibitem [{\citenamefont {Jim{\'e}nez}\ \emph {et~al.}(1993)\citenamefont {Jim{\'e}nez}, \citenamefont {Wray}, \citenamefont {Saffman},\ and\ \citenamefont {Rogallo}}]{Jimenez1993}%
  \BibitemOpen
  \bibfield  {author} {\bibinfo {author} {\bibfnamefont {J.}~\bibnamefont {Jim{\'e}nez}}, \bibinfo {author} {\bibfnamefont {A.~A.}\ \bibnamefont {Wray}}, \bibinfo {author} {\bibfnamefont {P.~G.}\ \bibnamefont {Saffman}},\ and\ \bibinfo {author} {\bibfnamefont {R.~S.}\ \bibnamefont {Rogallo}},\ }\href@noop {} {\bibfield  {journal} {\bibinfo  {journal} {Journal of Fluid Mechanics}\ }\textbf {\bibinfo {volume} {255}},\ \bibinfo {pages} {65} (\bibinfo {year} {1993})}\BibitemShut {NoStop}%
\bibitem [{\citenamefont {O{'}Connor}\ \emph {et~al.}(2016)\citenamefont {O{'}Connor}, \citenamefont {Day}, \citenamefont {Mandal},\ and\ \citenamefont {Revell}}]{OConnor2016}%
  \BibitemOpen
  \bibfield  {author} {\bibinfo {author} {\bibfnamefont {J.}~\bibnamefont {O{'}Connor}}, \bibinfo {author} {\bibfnamefont {P.}~\bibnamefont {Day}}, \bibinfo {author} {\bibfnamefont {P.}~\bibnamefont {Mandal}},\ and\ \bibinfo {author} {\bibfnamefont {A.}~\bibnamefont {Revell}},\ }\href {https://doi.org/10.1039/C6IB00009F} {\bibfield  {journal} {\bibinfo  {journal} {Integr. Biol.}\ }\textbf {\bibinfo {volume} {8}},\ \bibinfo {pages} {589} (\bibinfo {year} {2016})}\BibitemShut {NoStop}%
\bibitem [{\citenamefont {Anderson}(1995)}]{anderson1995computational}%
  \BibitemOpen
  \bibfield  {author} {\bibinfo {author} {\bibfnamefont {J.}~\bibnamefont {Anderson}},\ }\href {https://books.google.co.uk/books?id=phG\_QgAACAAJ} {\emph {\bibinfo {title} {Computational Fluid Dynamics: The Basics with Applications}}},\ McGraw-Hill International Editions: Mechanical Engineering\ (\bibinfo  {publisher} {McGraw-Hill},\ \bibinfo {year} {1995})\BibitemShut {NoStop}%
\bibitem [{\citenamefont {Hussain}(1986)}]{Hussain1986}%
  \BibitemOpen
  \bibfield  {author} {\bibinfo {author} {\bibfnamefont {A.~K. M.~F.}\ \bibnamefont {Hussain}},\ }\href@noop {} {\bibfield  {journal} {\bibinfo  {journal} {Journal of Fluid Mechanics}\ }\textbf {\bibinfo {volume} {173}},\ \bibinfo {pages} {303} (\bibinfo {year} {1986})}\BibitemShut {NoStop}%
\bibitem [{\citenamefont {Jeong}\ and\ \citenamefont {Hussain}(1995)}]{Jeong1995}%
  \BibitemOpen
  \bibfield  {author} {\bibinfo {author} {\bibfnamefont {J.}~\bibnamefont {Jeong}}\ and\ \bibinfo {author} {\bibfnamefont {F.}~\bibnamefont {Hussain}},\ }\href@noop {} {\bibfield  {journal} {\bibinfo  {journal} {Journal of Fluid Mechanics}\ }\textbf {\bibinfo {volume} {285}},\ \bibinfo {pages} {69} (\bibinfo {year} {1995})}\BibitemShut {NoStop}%
\bibitem [{\citenamefont {Lindner}\ \emph {et~al.}(2020)\citenamefont {Lindner}, \citenamefont {Devaux},\ and\ \citenamefont {Miskovic}}]{coherent_structures}%
  \BibitemOpen
  \bibfield  {author} {\bibinfo {author} {\bibfnamefont {G.}~\bibnamefont {Lindner}}, \bibinfo {author} {\bibfnamefont {Y.}~\bibnamefont {Devaux}},\ and\ \bibinfo {author} {\bibfnamefont {S.}~\bibnamefont {Miskovic}},\ }\href {https://doi.org/https://doi.org/10.1016/j.softx.2020.100604} {\bibfield  {journal} {\bibinfo  {journal} {SoftwareX}\ }\textbf {\bibinfo {volume} {12}},\ \bibinfo {pages} {100604} (\bibinfo {year} {2020})}\BibitemShut {NoStop}%
\bibitem [{\citenamefont {Shibata}\ and\ \citenamefont {Magara}(2011)}]{Shibata2011}%
  \BibitemOpen
  \bibfield  {author} {\bibinfo {author} {\bibfnamefont {K.}~\bibnamefont {Shibata}}\ and\ \bibinfo {author} {\bibfnamefont {T.}~\bibnamefont {Magara}},\ }\href {https://doi.org/10.12942/lrsp-2011-6} {\bibfield  {journal} {\bibinfo  {journal} {Living Reviews in Solar Physics}\ }\textbf {\bibinfo {volume} {8}},\ \bibinfo {pages} {6} (\bibinfo {year} {2011})}\BibitemShut {NoStop}%
\bibitem [{\citenamefont {Goldreich}\ and\ \citenamefont {Sridhar}(1995)}]{Goldreich1995}%
  \BibitemOpen
  \bibfield  {author} {\bibinfo {author} {\bibfnamefont {P.}~\bibnamefont {Goldreich}}\ and\ \bibinfo {author} {\bibfnamefont {S.}~\bibnamefont {Sridhar}},\ }\href@noop {} {\bibfield  {journal} {\bibinfo  {journal} {The Astrophysical Journal}\ }\textbf {\bibinfo {volume} {438}},\ \bibinfo {pages} {763} (\bibinfo {year} {1995})}\BibitemShut {NoStop}%
\bibitem [{\citenamefont {Mac~Low}\ and\ \citenamefont {Klessen}(2004)}]{MacLow2004}%
  \BibitemOpen
  \bibfield  {author} {\bibinfo {author} {\bibfnamefont {M.-M.}\ \bibnamefont {Mac~Low}}\ and\ \bibinfo {author} {\bibfnamefont {R.~S.}\ \bibnamefont {Klessen}},\ }\href@noop {} {\bibfield  {journal} {\bibinfo  {journal} {Reviews of Modern Physics}\ }\textbf {\bibinfo {volume} {76}},\ \bibinfo {pages} {125} (\bibinfo {year} {2004})}\BibitemShut {NoStop}%
\bibitem [{\citenamefont {Couston}\ \emph {et~al.}(2020)\citenamefont {Couston}, \citenamefont {Lecoanet}, \citenamefont {Favier},\ and\ \citenamefont {Le~Bars}}]{Couston2020}%
  \BibitemOpen
  \bibfield  {author} {\bibinfo {author} {\bibfnamefont {L.-A.}\ \bibnamefont {Couston}}, \bibinfo {author} {\bibfnamefont {D.}~\bibnamefont {Lecoanet}}, \bibinfo {author} {\bibfnamefont {B.}~\bibnamefont {Favier}},\ and\ \bibinfo {author} {\bibfnamefont {M.}~\bibnamefont {Le~Bars}},\ }\href {https://doi.org/10.1103/PhysRevResearch.2.023143} {\bibfield  {journal} {\bibinfo  {journal} {Phys. Rev. Res.}\ }\textbf {\bibinfo {volume} {2}},\ \bibinfo {pages} {023143} (\bibinfo {year} {2020})}\BibitemShut {NoStop}%
\bibitem [{\citenamefont {McWilliams}(1984)}]{McWilliams1984}%
  \BibitemOpen
  \bibfield  {author} {\bibinfo {author} {\bibfnamefont {J.~C.}\ \bibnamefont {McWilliams}},\ }\href@noop {} {\bibfield  {journal} {\bibinfo  {journal} {Journal of Fluid Mechanics}\ }\textbf {\bibinfo {volume} {146}},\ \bibinfo {pages} {21} (\bibinfo {year} {1984})}\BibitemShut {NoStop}%
\bibitem [{\citenamefont {Spalart}\ and\ \citenamefont {Allmaras}(1994)}]{Spalart1994}%
  \BibitemOpen
  \bibfield  {author} {\bibinfo {author} {\bibfnamefont {P.~R.}\ \bibnamefont {Spalart}}\ and\ \bibinfo {author} {\bibfnamefont {S.~R.}\ \bibnamefont {Allmaras}},\ }\href@noop {} {\bibfield  {journal} {\bibinfo  {journal} {La Recherche A{\'e}rospatiale}\ ,\ \bibinfo {pages} {5}} (\bibinfo {year} {1994})}\BibitemShut {NoStop}%
\bibitem [{\citenamefont {Smagorinsky}(1963)}]{Smagorinsky1963}%
  \BibitemOpen
  \bibfield  {author} {\bibinfo {author} {\bibfnamefont {J.}~\bibnamefont {Smagorinsky}},\ }\href@noop {} {\bibfield  {journal} {\bibinfo  {journal} {Monthly Weather Review}\ }\textbf {\bibinfo {volume} {91}},\ \bibinfo {pages} {99} (\bibinfo {year} {1963})}\BibitemShut {NoStop}%
\bibitem [{\citenamefont {Sadlo}\ \emph {et~al.}(2004)\citenamefont {Sadlo}, \citenamefont {Peikert},\ and\ \citenamefont {Parkinson}}]{cfd_turbine_design}%
  \BibitemOpen
  \bibfield  {author} {\bibinfo {author} {\bibfnamefont {F.}~\bibnamefont {Sadlo}}, \bibinfo {author} {\bibfnamefont {R.}~\bibnamefont {Peikert}},\ and\ \bibinfo {author} {\bibfnamefont {E.}~\bibnamefont {Parkinson}},\ }in\ \href {https://doi.org/10.1109/VISUAL.2004.128} {\emph {\bibinfo {booktitle} {IEEE Visualization 2004}}}\ (\bibinfo {year} {2004})\ pp.\ \bibinfo {pages} {179--186}\BibitemShut {NoStop}%
\bibitem [{\citenamefont {Yang}\ \emph {et~al.}(2023)\citenamefont {Yang}, \citenamefont {Li}, \citenamefont {Ji}, \citenamefont {Shi}, \citenamefont {Pu}, \citenamefont {Long},\ and\ \citenamefont {He}}]{cfd_noise_study}%
  \BibitemOpen
  \bibfield  {author} {\bibinfo {author} {\bibfnamefont {Q.}~\bibnamefont {Yang}}, \bibinfo {author} {\bibfnamefont {W.}~\bibnamefont {Li}}, \bibinfo {author} {\bibfnamefont {L.}~\bibnamefont {Ji}}, \bibinfo {author} {\bibfnamefont {W.}~\bibnamefont {Shi}}, \bibinfo {author} {\bibfnamefont {W.}~\bibnamefont {Pu}}, \bibinfo {author} {\bibfnamefont {Y.}~\bibnamefont {Long}},\ and\ \bibinfo {author} {\bibfnamefont {X.}~\bibnamefont {He}},\ }\href@noop {} {\bibfield  {journal} {\bibinfo  {journal} {Journal of Marine Science and Engineering}\ }\textbf {\bibinfo {volume} {11}},\ \bibinfo {pages} {2209} (\bibinfo {year} {2023})}\BibitemShut {NoStop}%
\bibitem [{\citenamefont {Vermeer}\ \emph {et~al.}(2003)\citenamefont {Vermeer}, \citenamefont {S{\o}rensen},\ and\ \citenamefont {Crespo}}]{Vermeer2003}%
  \BibitemOpen
  \bibfield  {author} {\bibinfo {author} {\bibfnamefont {L.~J.}\ \bibnamefont {Vermeer}}, \bibinfo {author} {\bibfnamefont {J.~N.}\ \bibnamefont {S{\o}rensen}},\ and\ \bibinfo {author} {\bibfnamefont {A.}~\bibnamefont {Crespo}},\ }\href@noop {} {\bibfield  {journal} {\bibinfo  {journal} {Progress in Aerospace Sciences}\ }\textbf {\bibinfo {volume} {39}},\ \bibinfo {pages} {467} (\bibinfo {year} {2003})}\BibitemShut {NoStop}%
\bibitem [{\citenamefont {Lighthill}(1952)}]{Lighthill1952}%
  \BibitemOpen
  \bibfield  {author} {\bibinfo {author} {\bibfnamefont {M.~J.}\ \bibnamefont {Lighthill}},\ }\href@noop {} {\bibfield  {journal} {\bibinfo  {journal} {Proceedings of the Royal Society of London. Series A}\ }\textbf {\bibinfo {volume} {211}},\ \bibinfo {pages} {564} (\bibinfo {year} {1952})}\BibitemShut {NoStop}%
\bibitem [{\citenamefont {Bearman}(1984)}]{Bearman1984}%
  \BibitemOpen
  \bibfield  {author} {\bibinfo {author} {\bibfnamefont {P.~W.}\ \bibnamefont {Bearman}},\ }\href@noop {} {\bibfield  {journal} {\bibinfo  {journal} {Annual Review of Fluid Mechanics}\ }\textbf {\bibinfo {volume} {16}},\ \bibinfo {pages} {195} (\bibinfo {year} {1984})}\BibitemShut {NoStop}%
\bibitem [{\citenamefont {Moin}\ and\ \citenamefont {Mahesh}(1998)}]{Moin1998}%
  \BibitemOpen
  \bibfield  {author} {\bibinfo {author} {\bibfnamefont {P.}~\bibnamefont {Moin}}\ and\ \bibinfo {author} {\bibfnamefont {K.}~\bibnamefont {Mahesh}},\ }\href@noop {} {\bibfield  {journal} {\bibinfo  {journal} {Annual Review of Fluid Mechanics}\ }\textbf {\bibinfo {volume} {30}},\ \bibinfo {pages} {539} (\bibinfo {year} {1998})}\BibitemShut {NoStop}%
\bibitem [{\citenamefont {Kaneda}\ \emph {et~al.}(2003)\citenamefont {Kaneda}, \citenamefont {Ishihara}, \citenamefont {Yokokawa}, \citenamefont {Itakura},\ and\ \citenamefont {Uno}}]{Kaneda2003}%
  \BibitemOpen
  \bibfield  {author} {\bibinfo {author} {\bibfnamefont {Y.}~\bibnamefont {Kaneda}}, \bibinfo {author} {\bibfnamefont {T.}~\bibnamefont {Ishihara}}, \bibinfo {author} {\bibfnamefont {M.}~\bibnamefont {Yokokawa}}, \bibinfo {author} {\bibfnamefont {K.}~\bibnamefont {Itakura}},\ and\ \bibinfo {author} {\bibfnamefont {A.}~\bibnamefont {Uno}},\ }\href@noop {} {\bibfield  {journal} {\bibinfo  {journal} {Physics of Fluids}\ }\textbf {\bibinfo {volume} {15}},\ \bibinfo {pages} {L21} (\bibinfo {year} {2003})}\BibitemShut {NoStop}%
\bibitem [{\citenamefont {Lee}\ and\ \citenamefont {Moser}(2015)}]{Lee2015}%
  \BibitemOpen
  \bibfield  {author} {\bibinfo {author} {\bibfnamefont {M.}~\bibnamefont {Lee}}\ and\ \bibinfo {author} {\bibfnamefont {R.~D.}\ \bibnamefont {Moser}},\ }\href@noop {} {\bibfield  {journal} {\bibinfo  {journal} {Journal of Fluid Mechanics}\ }\textbf {\bibinfo {volume} {774}},\ \bibinfo {pages} {395} (\bibinfo {year} {2015})}\BibitemShut {NoStop}%
\bibitem [{\citenamefont {Harrow}\ \emph {et~al.}(2009)\citenamefont {Harrow}, \citenamefont {Hassidim},\ and\ \citenamefont {Lloyd}}]{harrow2009quantum}%
  \BibitemOpen
  \bibfield  {author} {\bibinfo {author} {\bibfnamefont {A.~W.}\ \bibnamefont {Harrow}}, \bibinfo {author} {\bibfnamefont {A.}~\bibnamefont {Hassidim}},\ and\ \bibinfo {author} {\bibfnamefont {S.}~\bibnamefont {Lloyd}},\ }\href {https://journals.aps.org/prl/abstract/10.1103/PhysRevLett.103.150502} {\bibfield  {journal} {\bibinfo  {journal} {Physical review letters}\ }\textbf {\bibinfo {volume} {103}},\ \bibinfo {pages} {150502} (\bibinfo {year} {2009})}\BibitemShut {NoStop}%
\bibitem [{\citenamefont {Dalzell}\ \emph {et~al.}(2023)\citenamefont {Dalzell}, \citenamefont {McArdle}, \citenamefont {Berta}, \citenamefont {Bienias}, \citenamefont {Chen}, \citenamefont {Gilyén}, \citenamefont {Hann}, \citenamefont {Kastoryano}, \citenamefont {Khabiboulline}, \citenamefont {Kubica}, \citenamefont {Salton}, \citenamefont {Wang},\ and\ \citenamefont {Brandão}}]{algorithms_survey}%
  \BibitemOpen
  \bibfield  {author} {\bibinfo {author} {\bibfnamefont {A.~M.}\ \bibnamefont {Dalzell}}, \bibinfo {author} {\bibfnamefont {S.}~\bibnamefont {McArdle}}, \bibinfo {author} {\bibfnamefont {M.}~\bibnamefont {Berta}}, \bibinfo {author} {\bibfnamefont {P.}~\bibnamefont {Bienias}}, \bibinfo {author} {\bibfnamefont {C.-F.}\ \bibnamefont {Chen}}, \bibinfo {author} {\bibfnamefont {A.}~\bibnamefont {Gilyén}}, \bibinfo {author} {\bibfnamefont {C.~T.}\ \bibnamefont {Hann}}, \bibinfo {author} {\bibfnamefont {M.~J.}\ \bibnamefont {Kastoryano}}, \bibinfo {author} {\bibfnamefont {E.~T.}\ \bibnamefont {Khabiboulline}}, \bibinfo {author} {\bibfnamefont {A.}~\bibnamefont {Kubica}}, \bibinfo {author} {\bibfnamefont {G.}~\bibnamefont {Salton}}, \bibinfo {author} {\bibfnamefont {S.}~\bibnamefont {Wang}},\ and\ \bibinfo {author} {\bibfnamefont {F.~G. S.~L.}\ \bibnamefont {Brandão}},\ }\href@noop {} {\bibinfo {title} {Quantum algorithms: A survey of applications and end-to-end complexities}} (\bibinfo {year} {2023}),\ \Eprint
  {https://arxiv.org/abs/2310.03011} {arXiv:2310.03011 [quant-ph]} \BibitemShut {NoStop}%
\bibitem [{\citenamefont {Cao}\ \emph {et~al.}(2013)\citenamefont {Cao}, \citenamefont {Papageorgiou}, \citenamefont {Petras}, \citenamefont {Traub},\ and\ \citenamefont {Kais}}]{Cao_2013}%
  \BibitemOpen
  \bibfield  {author} {\bibinfo {author} {\bibfnamefont {Y.}~\bibnamefont {Cao}}, \bibinfo {author} {\bibfnamefont {A.}~\bibnamefont {Papageorgiou}}, \bibinfo {author} {\bibfnamefont {I.}~\bibnamefont {Petras}}, \bibinfo {author} {\bibfnamefont {J.}~\bibnamefont {Traub}},\ and\ \bibinfo {author} {\bibfnamefont {S.}~\bibnamefont {Kais}},\ }\href {https://doi.org/10.1088/1367-2630/15/1/013021} {\bibfield  {journal} {\bibinfo  {journal} {New Journal of Physics}\ }\textbf {\bibinfo {volume} {15}},\ \bibinfo {pages} {013021} (\bibinfo {year} {2013})}\BibitemShut {NoStop}%
\bibitem [{\citenamefont {Montanaro}\ and\ \citenamefont {Pallister}(2016)}]{Montanaro2016}%
  \BibitemOpen
  \bibfield  {author} {\bibinfo {author} {\bibfnamefont {A.}~\bibnamefont {Montanaro}}\ and\ \bibinfo {author} {\bibfnamefont {S.}~\bibnamefont {Pallister}},\ }\href {https://doi.org/10.1103/PhysRevA.93.032324} {\bibfield  {journal} {\bibinfo  {journal} {Phys. Rev. A}\ }\textbf {\bibinfo {volume} {93}},\ \bibinfo {pages} {032324} (\bibinfo {year} {2016})}\BibitemShut {NoStop}%
\bibitem [{\citenamefont {Xin}\ \emph {et~al.}(2020)\citenamefont {Xin}, \citenamefont {Wei}, \citenamefont {Cui}, \citenamefont {Xiao}, \citenamefont {Arrazola}, \citenamefont {Lamata}, \citenamefont {Kong}, \citenamefont {Lu}, \citenamefont {Solano},\ and\ \citenamefont {Long}}]{Xin2020}%
  \BibitemOpen
  \bibfield  {author} {\bibinfo {author} {\bibfnamefont {T.}~\bibnamefont {Xin}}, \bibinfo {author} {\bibfnamefont {S.}~\bibnamefont {Wei}}, \bibinfo {author} {\bibfnamefont {J.}~\bibnamefont {Cui}}, \bibinfo {author} {\bibfnamefont {J.}~\bibnamefont {Xiao}}, \bibinfo {author} {\bibfnamefont {I.~n.}\ \bibnamefont {Arrazola}}, \bibinfo {author} {\bibfnamefont {L.}~\bibnamefont {Lamata}}, \bibinfo {author} {\bibfnamefont {X.}~\bibnamefont {Kong}}, \bibinfo {author} {\bibfnamefont {D.}~\bibnamefont {Lu}}, \bibinfo {author} {\bibfnamefont {E.}~\bibnamefont {Solano}},\ and\ \bibinfo {author} {\bibfnamefont {G.}~\bibnamefont {Long}},\ }\href {https://doi.org/10.1103/PhysRevA.101.032307} {\bibfield  {journal} {\bibinfo  {journal} {Phys. Rev. A}\ }\textbf {\bibinfo {volume} {101}},\ \bibinfo {pages} {032307} (\bibinfo {year} {2020})}\BibitemShut {NoStop}%
\bibitem [{\citenamefont {Martin}\ \emph {et~al.}(2023)\citenamefont {Martin}, \citenamefont {Ibarrondo},\ and\ \citenamefont {Sanz}}]{MartinSanz2023}%
  \BibitemOpen
  \bibfield  {author} {\bibinfo {author} {\bibfnamefont {A.}~\bibnamefont {Martin}}, \bibinfo {author} {\bibfnamefont {R.}~\bibnamefont {Ibarrondo}},\ and\ \bibinfo {author} {\bibfnamefont {M.}~\bibnamefont {Sanz}},\ }\href {https://doi.org/10.1103/PhysRevApplied.19.064056} {\bibfield  {journal} {\bibinfo  {journal} {Phys. Rev. Appl.}\ }\textbf {\bibinfo {volume} {19}},\ \bibinfo {pages} {064056} (\bibinfo {year} {2023})}\BibitemShut {NoStop}%
\bibitem [{\citenamefont {Williams}\ \emph {et~al.}(2023)\citenamefont {Williams}, \citenamefont {Paine}, \citenamefont {Wu}, \citenamefont {Elfving},\ and\ \citenamefont {Kyriienko}}]{Williams2023}%
  \BibitemOpen
  \bibfield  {author} {\bibinfo {author} {\bibfnamefont {C.~A.}\ \bibnamefont {Williams}}, \bibinfo {author} {\bibfnamefont {A.~E.}\ \bibnamefont {Paine}}, \bibinfo {author} {\bibfnamefont {H.-Y.}\ \bibnamefont {Wu}}, \bibinfo {author} {\bibfnamefont {V.~E.}\ \bibnamefont {Elfving}},\ and\ \bibinfo {author} {\bibfnamefont {O.}~\bibnamefont {Kyriienko}},\ }\href@noop {} {\bibinfo {title} {Quantum chebyshev transform: Mapping, embedding, learning and sampling distributions}} (\bibinfo {year} {2023}),\ \Eprint {https://arxiv.org/abs/2306.17026} {arXiv:2306.17026 [quant-ph]} \BibitemShut {NoStop}%
\bibitem [{\citenamefont {Kyriienko}\ \emph {et~al.}(2024)\citenamefont {Kyriienko}, \citenamefont {Paine},\ and\ \citenamefont {Elfving}}]{Kyriienko2024protocols}%
  \BibitemOpen
  \bibfield  {author} {\bibinfo {author} {\bibfnamefont {O.}~\bibnamefont {Kyriienko}}, \bibinfo {author} {\bibfnamefont {A.~E.}\ \bibnamefont {Paine}},\ and\ \bibinfo {author} {\bibfnamefont {V.~E.}\ \bibnamefont {Elfving}},\ }\href {https://doi.org/10.1103/PhysRevResearch.6.033291} {\bibfield  {journal} {\bibinfo  {journal} {Phys. Rev. Res.}\ }\textbf {\bibinfo {volume} {6}},\ \bibinfo {pages} {033291} (\bibinfo {year} {2024})}\BibitemShut {NoStop}%
\bibitem [{\citenamefont {Lin}\ and\ \citenamefont {Tong}(2020)}]{Lin2020optimalpolynomial}%
  \BibitemOpen
  \bibfield  {author} {\bibinfo {author} {\bibfnamefont {L.}~\bibnamefont {Lin}}\ and\ \bibinfo {author} {\bibfnamefont {Y.}~\bibnamefont {Tong}},\ }\href {https://doi.org/10.22331/q-2020-11-11-361} {\bibfield  {journal} {\bibinfo  {journal} {{Quantum}}\ }\textbf {\bibinfo {volume} {4}},\ \bibinfo {pages} {361} (\bibinfo {year} {2020})}\BibitemShut {NoStop}%
\bibitem [{\citenamefont {Linden}\ \emph {et~al.}(2022)\citenamefont {Linden}, \citenamefont {Montanaro},\ and\ \citenamefont {Shao}}]{Linden2022quantumvsclassical}%
  \BibitemOpen
  \bibfield  {author} {\bibinfo {author} {\bibfnamefont {N.}~\bibnamefont {Linden}}, \bibinfo {author} {\bibfnamefont {A.}~\bibnamefont {Montanaro}},\ and\ \bibinfo {author} {\bibfnamefont {C.}~\bibnamefont {Shao}},\ }\href {https://doi.org/10.1007/s00220-022-04442-6} {\bibfield  {journal} {\bibinfo  {journal} {Communications in Mathematical Physics}\ }\textbf {\bibinfo {volume} {395}},\ \bibinfo {pages} {601} (\bibinfo {year} {2022})}\BibitemShut {NoStop}%
\bibitem [{\citenamefont {Krovi}(2023)}]{Krovi2023improvedquantum}%
  \BibitemOpen
  \bibfield  {author} {\bibinfo {author} {\bibfnamefont {H.}~\bibnamefont {Krovi}},\ }\href {https://doi.org/10.22331/q-2023-02-02-913} {\bibfield  {journal} {\bibinfo  {journal} {{Quantum}}\ }\textbf {\bibinfo {volume} {7}},\ \bibinfo {pages} {913} (\bibinfo {year} {2023})}\BibitemShut {NoStop}%
\bibitem [{\citenamefont {Costa}\ \emph {et~al.}(2023)\citenamefont {Costa}, \citenamefont {Schleich}, \citenamefont {Morales},\ and\ \citenamefont {Berry}}]{costa2023improving}%
  \BibitemOpen
  \bibfield  {author} {\bibinfo {author} {\bibfnamefont {P.~C.~S.}\ \bibnamefont {Costa}}, \bibinfo {author} {\bibfnamefont {P.}~\bibnamefont {Schleich}}, \bibinfo {author} {\bibfnamefont {M.~E.~S.}\ \bibnamefont {Morales}},\ and\ \bibinfo {author} {\bibfnamefont {D.~W.}\ \bibnamefont {Berry}},\ }\href@noop {} {\bibinfo {title} {Further improving quantum algorithms for nonlinear differential equations via higher-order methods and rescaling}} (\bibinfo {year} {2023}),\ \Eprint {https://arxiv.org/abs/2312.09518} {arXiv:2312.09518 [quant-ph]} \BibitemShut {NoStop}%
\bibitem [{\citenamefont {Childs}\ \emph {et~al.}(2017)\citenamefont {Childs}, \citenamefont {Kothari},\ and\ \citenamefont {Somma}}]{childs2017quantum}%
  \BibitemOpen
  \bibfield  {author} {\bibinfo {author} {\bibfnamefont {A.~M.}\ \bibnamefont {Childs}}, \bibinfo {author} {\bibfnamefont {R.}~\bibnamefont {Kothari}},\ and\ \bibinfo {author} {\bibfnamefont {R.~D.}\ \bibnamefont {Somma}},\ }\href@noop {} {\bibfield  {journal} {\bibinfo  {journal} {SIAM Journal on Computing}\ }\textbf {\bibinfo {volume} {46}},\ \bibinfo {pages} {1920} (\bibinfo {year} {2017})}\BibitemShut {NoStop}%
\bibitem [{\citenamefont {Childs}\ and\ \citenamefont {Wiebe}(2012)}]{Childs2012}%
  \BibitemOpen
  \bibfield  {author} {\bibinfo {author} {\bibfnamefont {A.~M.}\ \bibnamefont {Childs}}\ and\ \bibinfo {author} {\bibfnamefont {N.}~\bibnamefont {Wiebe}},\ }\href@noop {} {\bibfield  {journal} {\bibinfo  {journal} {Quantum Info. Comput.}\ }\textbf {\bibinfo {volume} {12}},\ \bibinfo {pages} {901–924} (\bibinfo {year} {2012})}\BibitemShut {NoStop}%
\bibitem [{\citenamefont {Berry}\ \emph {et~al.}(2017)\citenamefont {Berry}, \citenamefont {Childs}, \citenamefont {Ostrander},\ and\ \citenamefont {Wang}}]{Berry2017}%
  \BibitemOpen
  \bibfield  {author} {\bibinfo {author} {\bibfnamefont {D.~W.}\ \bibnamefont {Berry}}, \bibinfo {author} {\bibfnamefont {A.~M.}\ \bibnamefont {Childs}}, \bibinfo {author} {\bibfnamefont {A.}~\bibnamefont {Ostrander}},\ and\ \bibinfo {author} {\bibfnamefont {G.}~\bibnamefont {Wang}},\ }\href {https://doi.org/10.1007/s00220-017-3002-y} {\bibfield  {journal} {\bibinfo  {journal} {Communications in Mathematical Physics}\ }\textbf {\bibinfo {volume} {356}},\ \bibinfo {pages} {1057} (\bibinfo {year} {2017})}\BibitemShut {NoStop}%
\bibitem [{\citenamefont {Jennings}\ \emph {et~al.}(2023)\citenamefont {Jennings}, \citenamefont {Lostaglio}, \citenamefont {Pallister}, \citenamefont {Sornborger},\ and\ \citenamefont {Subaşı}}]{jennings2023efficient}%
  \BibitemOpen
  \bibfield  {author} {\bibinfo {author} {\bibfnamefont {D.}~\bibnamefont {Jennings}}, \bibinfo {author} {\bibfnamefont {M.}~\bibnamefont {Lostaglio}}, \bibinfo {author} {\bibfnamefont {S.}~\bibnamefont {Pallister}}, \bibinfo {author} {\bibfnamefont {A.~T.}\ \bibnamefont {Sornborger}},\ and\ \bibinfo {author} {\bibfnamefont {Y.}~\bibnamefont {Subaşı}},\ }\href@noop {} {\bibinfo {title} {Efficient quantum linear solver algorithm with detailed running costs}} (\bibinfo {year} {2023}),\ \Eprint {https://arxiv.org/abs/2305.11352} {arXiv:2305.11352 [quant-ph]} \BibitemShut {NoStop}%
\bibitem [{\citenamefont {Gribling}\ \emph {et~al.}(2024)\citenamefont {Gribling}, \citenamefont {Kerenidis},\ and\ \citenamefont {Szil\'{a}gyi}}]{Gribling2024}%
  \BibitemOpen
  \bibfield  {author} {\bibinfo {author} {\bibfnamefont {S.}~\bibnamefont {Gribling}}, \bibinfo {author} {\bibfnamefont {I.}~\bibnamefont {Kerenidis}},\ and\ \bibinfo {author} {\bibfnamefont {D.}~\bibnamefont {Szil\'{a}gyi}},\ }\bibfield  {journal} {\bibinfo  {journal} {ACM Transactions on Quantum Computing}\ }\textbf {\bibinfo {volume} {5}},\ \href {https://doi.org/10.1145/3649320} {10.1145/3649320} (\bibinfo {year} {2024})\BibitemShut {NoStop}%
\bibitem [{\citenamefont {Jin}\ \emph {et~al.}(2023)\citenamefont {Jin}, \citenamefont {Liu},\ and\ \citenamefont {Yu}}]{jin2022quantum}%
  \BibitemOpen
  \bibfield  {author} {\bibinfo {author} {\bibfnamefont {S.}~\bibnamefont {Jin}}, \bibinfo {author} {\bibfnamefont {N.}~\bibnamefont {Liu}},\ and\ \bibinfo {author} {\bibfnamefont {Y.}~\bibnamefont {Yu}},\ }\href {https://doi.org/10.1103/PhysRevA.108.032603} {\bibfield  {journal} {\bibinfo  {journal} {Phys. Rev. A}\ }\textbf {\bibinfo {volume} {108}},\ \bibinfo {pages} {032603} (\bibinfo {year} {2023})}\BibitemShut {NoStop}%
\bibitem [{\citenamefont {Jin}\ \emph {et~al.}(2022)\citenamefont {Jin}, \citenamefont {Liu},\ and\ \citenamefont {Yu}}]{jin2022quantumEXT}%
  \BibitemOpen
  \bibfield  {author} {\bibinfo {author} {\bibfnamefont {S.}~\bibnamefont {Jin}}, \bibinfo {author} {\bibfnamefont {N.}~\bibnamefont {Liu}},\ and\ \bibinfo {author} {\bibfnamefont {Y.}~\bibnamefont {Yu}},\ }\href@noop {} {\bibinfo {title} {Quantum simulation of partial differential equations via schrodingerisation: technical details}} (\bibinfo {year} {2022}),\ \Eprint {https://arxiv.org/abs/2212.14703} {arXiv:2212.14703 [quant-ph]} \BibitemShut {NoStop}%
\bibitem [{\citenamefont {Jin}\ and\ \citenamefont {Liu}(2023{\natexlab{a}})}]{jin2023analog}%
  \BibitemOpen
  \bibfield  {author} {\bibinfo {author} {\bibfnamefont {S.}~\bibnamefont {Jin}}\ and\ \bibinfo {author} {\bibfnamefont {N.}~\bibnamefont {Liu}},\ }\href@noop {} {\bibinfo {title} {Analog quantum simulation of partial differential equations}} (\bibinfo {year} {2023}{\natexlab{a}}),\ \Eprint {https://arxiv.org/abs/2308.00646} {arXiv:2308.00646 [quant-ph]} \BibitemShut {NoStop}%
\bibitem [{\citenamefont {Jin}\ and\ \citenamefont {Liu}(2023{\natexlab{b}})}]{jin2023quantum}%
  \BibitemOpen
  \bibfield  {author} {\bibinfo {author} {\bibfnamefont {S.}~\bibnamefont {Jin}}\ and\ \bibinfo {author} {\bibfnamefont {N.}~\bibnamefont {Liu}},\ }\href@noop {} {\bibinfo {title} {Quantum simulation of discrete linear dynamical systems and simple iterative methods in linear algebra via schrodingerisation}} (\bibinfo {year} {2023}{\natexlab{b}}),\ \Eprint {https://arxiv.org/abs/2304.02865} {arXiv:2304.02865 [quant-ph]} \BibitemShut {NoStop}%
\bibitem [{\citenamefont {Hu}\ \emph {et~al.}(2024)\citenamefont {Hu}, \citenamefont {Jin}, \citenamefont {Liu},\ and\ \citenamefont {Zhang}}]{hu2024quantum}%
  \BibitemOpen
  \bibfield  {author} {\bibinfo {author} {\bibfnamefont {J.}~\bibnamefont {Hu}}, \bibinfo {author} {\bibfnamefont {S.}~\bibnamefont {Jin}}, \bibinfo {author} {\bibfnamefont {N.}~\bibnamefont {Liu}},\ and\ \bibinfo {author} {\bibfnamefont {L.}~\bibnamefont {Zhang}},\ }\href@noop {} {\bibinfo {title} {Quantum circuits for partial differential equations via schr\"odingerisation}} (\bibinfo {year} {2024}),\ \Eprint {https://arxiv.org/abs/2403.10032} {arXiv:2403.10032 [quant-ph]} \BibitemShut {NoStop}%
\bibitem [{\citenamefont {Liu}\ \emph {et~al.}(2024)\citenamefont {Liu}, \citenamefont {Ortiz},\ and\ \citenamefont {Cirak}}]{LiuCirak2024}%
  \BibitemOpen
  \bibfield  {author} {\bibinfo {author} {\bibfnamefont {B.}~\bibnamefont {Liu}}, \bibinfo {author} {\bibfnamefont {M.}~\bibnamefont {Ortiz}},\ and\ \bibinfo {author} {\bibfnamefont {F.}~\bibnamefont {Cirak}},\ }\href {https://doi.org/https://doi.org/10.1016/j.cma.2024.117403} {\bibfield  {journal} {\bibinfo  {journal} {Computer Methods in Applied Mechanics and Engineering}\ }\textbf {\bibinfo {volume} {432}},\ \bibinfo {pages} {117403} (\bibinfo {year} {2024})}\BibitemShut {NoStop}%
\bibitem [{\citenamefont {Lubasch}\ \emph {et~al.}(2025)\citenamefont {Lubasch}, \citenamefont {Kikuchi}, \citenamefont {Wright},\ and\ \citenamefont {Keever}}]{lubasch2025fourier}%
  \BibitemOpen
  \bibfield  {author} {\bibinfo {author} {\bibfnamefont {M.}~\bibnamefont {Lubasch}}, \bibinfo {author} {\bibfnamefont {Y.}~\bibnamefont {Kikuchi}}, \bibinfo {author} {\bibfnamefont {L.}~\bibnamefont {Wright}},\ and\ \bibinfo {author} {\bibfnamefont {C.~M.}\ \bibnamefont {Keever}},\ }\href {https://arxiv.org/abs/2505.16895} {\bibinfo {title} {Quantum circuits for partial differential equations in fourier space}} (\bibinfo {year} {2025}),\ \Eprint {https://arxiv.org/abs/2505.16895} {arXiv:2505.16895 [quant-ph]} \BibitemShut {NoStop}%
\bibitem [{\citenamefont {Devereux}\ and\ \citenamefont {Datta}(2025)}]{devereux2025drift}%
  \BibitemOpen
  \bibfield  {author} {\bibinfo {author} {\bibfnamefont {E.}~\bibnamefont {Devereux}}\ and\ \bibinfo {author} {\bibfnamefont {A.}~\bibnamefont {Datta}},\ }\href {https://arxiv.org/abs/2505.21221} {\bibinfo {title} {Quantum algorithms for solving a drift-diffusion equation}} (\bibinfo {year} {2025}),\ \Eprint {https://arxiv.org/abs/2505.21221} {arXiv:2505.21221 [quant-ph]} \BibitemShut {NoStop}%
\bibitem [{\citenamefont {Suba{\c{s}}{\i}}\ \emph {et~al.}(2019)\citenamefont {Suba{\c{s}}{\i}}, \citenamefont {Somma},\ and\ \citenamefont {Orsucci}}]{subacsi2019quantum}%
  \BibitemOpen
  \bibfield  {author} {\bibinfo {author} {\bibfnamefont {Y.}~\bibnamefont {Suba{\c{s}}{\i}}}, \bibinfo {author} {\bibfnamefont {R.~D.}\ \bibnamefont {Somma}},\ and\ \bibinfo {author} {\bibfnamefont {D.}~\bibnamefont {Orsucci}},\ }\href@noop {} {\bibfield  {journal} {\bibinfo  {journal} {Physical Review Letters}\ }\textbf {\bibinfo {volume} {122}},\ \bibinfo {pages} {060504} (\bibinfo {year} {2019})}\BibitemShut {NoStop}%
\bibitem [{\citenamefont {Costa}\ \emph {et~al.}(2022)\citenamefont {Costa}, \citenamefont {An}, \citenamefont {Sanders}, \citenamefont {Su}, \citenamefont {Babbush},\ and\ \citenamefont {Berry}}]{Costa2022}%
  \BibitemOpen
  \bibfield  {author} {\bibinfo {author} {\bibfnamefont {P.~C.}\ \bibnamefont {Costa}}, \bibinfo {author} {\bibfnamefont {D.}~\bibnamefont {An}}, \bibinfo {author} {\bibfnamefont {Y.~R.}\ \bibnamefont {Sanders}}, \bibinfo {author} {\bibfnamefont {Y.}~\bibnamefont {Su}}, \bibinfo {author} {\bibfnamefont {R.}~\bibnamefont {Babbush}},\ and\ \bibinfo {author} {\bibfnamefont {D.~W.}\ \bibnamefont {Berry}},\ }\href {https://doi.org/10.1103/PRXQuantum.3.040303} {\bibfield  {journal} {\bibinfo  {journal} {PRX Quantum}\ }\textbf {\bibinfo {volume} {3}},\ \bibinfo {pages} {040303} (\bibinfo {year} {2022})}\BibitemShut {NoStop}%
\bibitem [{\citenamefont {An}\ and\ \citenamefont {Lin}(2022)}]{DongLin2022}%
  \BibitemOpen
  \bibfield  {author} {\bibinfo {author} {\bibfnamefont {D.}~\bibnamefont {An}}\ and\ \bibinfo {author} {\bibfnamefont {L.}~\bibnamefont {Lin}},\ }\bibfield  {journal} {\bibinfo  {journal} {ACM Transactions on Quantum Computing}\ }\textbf {\bibinfo {volume} {3}},\ \href {https://doi.org/10.1145/3498331} {10.1145/3498331} (\bibinfo {year} {2022})\BibitemShut {NoStop}%
\bibitem [{\citenamefont {Williams}\ \emph {et~al.}(2024{\natexlab{a}})\citenamefont {Williams}, \citenamefont {Gentile}, \citenamefont {Elfving}, \citenamefont {Berger},\ and\ \citenamefont {Kyriienko}}]{williams2024iterative}%
  \BibitemOpen
  \bibfield  {author} {\bibinfo {author} {\bibfnamefont {C.~A.}\ \bibnamefont {Williams}}, \bibinfo {author} {\bibfnamefont {A.~A.}\ \bibnamefont {Gentile}}, \bibinfo {author} {\bibfnamefont {V.~E.}\ \bibnamefont {Elfving}}, \bibinfo {author} {\bibfnamefont {D.}~\bibnamefont {Berger}},\ and\ \bibinfo {author} {\bibfnamefont {O.}~\bibnamefont {Kyriienko}},\ }\href {https://arxiv.org/abs/2404.08605} {\bibinfo {title} {Quantum iterative methods for solving differential equations with application to computational fluid dynamics}} (\bibinfo {year} {2024}{\natexlab{a}}),\ \Eprint {https://arxiv.org/abs/2404.08605} {arXiv:2404.08605 [quant-ph]} \BibitemShut {NoStop}%
\bibitem [{\citenamefont {Williams}\ \emph {et~al.}(2024{\natexlab{b}})\citenamefont {Williams}, \citenamefont {Gentile}, \citenamefont {Elfving}, \citenamefont {Berger},\ and\ \citenamefont {Kyriienko}}]{quantum_iterative_solver}%
  \BibitemOpen
  \bibfield  {author} {\bibinfo {author} {\bibfnamefont {C.~A.}\ \bibnamefont {Williams}}, \bibinfo {author} {\bibfnamefont {A.~A.}\ \bibnamefont {Gentile}}, \bibinfo {author} {\bibfnamefont {V.~E.}\ \bibnamefont {Elfving}}, \bibinfo {author} {\bibfnamefont {D.}~\bibnamefont {Berger}},\ and\ \bibinfo {author} {\bibfnamefont {O.}~\bibnamefont {Kyriienko}},\ }\href {https://arxiv.org/abs/2404.08605} {\bibinfo {title} {Quantum iterative methods for solving differential equations with application to computational fluid dynamics}} (\bibinfo {year} {2024}{\natexlab{b}}),\ \Eprint {https://arxiv.org/abs/2404.08605} {arXiv:2404.08605 [quant-ph]} \BibitemShut {NoStop}%
\bibitem [{\citenamefont {Raisuddin}\ and\ \citenamefont {De}(2024)}]{time_marching_multigrid}%
  \BibitemOpen
  \bibfield  {author} {\bibinfo {author} {\bibfnamefont {O.~M.}\ \bibnamefont {Raisuddin}}\ and\ \bibinfo {author} {\bibfnamefont {S.}~\bibnamefont {De}},\ }\href@noop {} {\bibinfo {title} {Quantum multigrid algorithm for finite element problems}} (\bibinfo {year} {2024}),\ \Eprint {https://arxiv.org/abs/2404.07466} {arXiv:2404.07466 [quant-ph]} \BibitemShut {NoStop}%
\bibitem [{\citenamefont {Jaksch}(2022)}]{quantum_multigrid}%
  \BibitemOpen
  \bibfield  {author} {\bibinfo {author} {\bibfnamefont {P.}~\bibnamefont {Jaksch}},\ }\href@noop {} {\bibinfo {title} {Implementation of a digitally encoded multigrid algorithm on a quantum computer}} (\bibinfo {year} {2022}),\ \Eprint {https://arxiv.org/abs/2201.04513} {arXiv:2201.04513 [quant-ph]} \BibitemShut {NoStop}%
\bibitem [{\citenamefont {Liu}\ \emph {et~al.}(2021)\citenamefont {Liu}, \citenamefont {Øie Kolden}, \citenamefont {Krovi}, \citenamefont {Loureiro}, \citenamefont {Trivisa},\ and\ \citenamefont {Childs}}]{JPLiu2021}%
  \BibitemOpen
  \bibfield  {author} {\bibinfo {author} {\bibfnamefont {J.-P.}\ \bibnamefont {Liu}}, \bibinfo {author} {\bibfnamefont {H.}~\bibnamefont {Øie Kolden}}, \bibinfo {author} {\bibfnamefont {H.~K.}\ \bibnamefont {Krovi}}, \bibinfo {author} {\bibfnamefont {N.~F.}\ \bibnamefont {Loureiro}}, \bibinfo {author} {\bibfnamefont {K.}~\bibnamefont {Trivisa}},\ and\ \bibinfo {author} {\bibfnamefont {A.~M.}\ \bibnamefont {Childs}},\ }\href {https://doi.org/10.1073/pnas.2026805118} {\bibfield  {journal} {\bibinfo  {journal} {Proceedings of the National Academy of Sciences}\ }\textbf {\bibinfo {volume} {118}},\ \bibinfo {pages} {e2026805118} (\bibinfo {year} {2021})},\ \Eprint {https://arxiv.org/abs/https://www.pnas.org/doi/pdf/10.1073/pnas.2026805118} {https://www.pnas.org/doi/pdf/10.1073/pnas.2026805118} \BibitemShut {NoStop}%
\bibitem [{\citenamefont {Tanaka}\ \emph {et~al.}(2023)\citenamefont {Tanaka}, \citenamefont {Ito},\ and\ \citenamefont {Fujii}}]{tanaka2023carleman}%
  \BibitemOpen
  \bibfield  {author} {\bibinfo {author} {\bibfnamefont {Y.}~\bibnamefont {Tanaka}}, \bibinfo {author} {\bibfnamefont {Y.}~\bibnamefont {Ito}},\ and\ \bibinfo {author} {\bibfnamefont {K.}~\bibnamefont {Fujii}},\ }\href {https://doi.org/10.1038/s41598-023-31009-9} {\bibfield  {journal} {\bibinfo  {journal} {Scientific Reports}\ }\textbf {\bibinfo {volume} {13}},\ \bibinfo {pages} {4531} (\bibinfo {year} {2023})}\BibitemShut {NoStop}%
\bibitem [{\citenamefont {Ingelmann}\ \emph {et~al.}(2023)\citenamefont {Ingelmann}, \citenamefont {Bharadwaj}, \citenamefont {Pfeffer}, \citenamefont {Sreenivasan},\ and\ \citenamefont {Schumacher}}]{ingelmann2023quantum}%
  \BibitemOpen
  \bibfield  {author} {\bibinfo {author} {\bibfnamefont {J.}~\bibnamefont {Ingelmann}}, \bibinfo {author} {\bibfnamefont {S.~S.}\ \bibnamefont {Bharadwaj}}, \bibinfo {author} {\bibfnamefont {P.}~\bibnamefont {Pfeffer}}, \bibinfo {author} {\bibfnamefont {K.~R.}\ \bibnamefont {Sreenivasan}},\ and\ \bibinfo {author} {\bibfnamefont {J.}~\bibnamefont {Schumacher}},\ }\href@noop {} {\bibinfo {title} {Two quantum algorithms for solving the one-dimensional advection-diffusion equation}} (\bibinfo {year} {2023}),\ \Eprint {https://arxiv.org/abs/2401.00326} {arXiv:2401.00326 [physics.flu-dyn]} \BibitemShut {NoStop}%
\bibitem [{\citenamefont {Wu}\ \emph {et~al.}(2024)\citenamefont {Wu}, \citenamefont {Wang},\ and\ \citenamefont {Li}}]{wu2024quantum}%
  \BibitemOpen
  \bibfield  {author} {\bibinfo {author} {\bibfnamefont {H.-C.}\ \bibnamefont {Wu}}, \bibinfo {author} {\bibfnamefont {J.}~\bibnamefont {Wang}},\ and\ \bibinfo {author} {\bibfnamefont {X.}~\bibnamefont {Li}},\ }\href@noop {} {\bibfield  {journal} {\bibinfo  {journal} {arXiv preprint arXiv:2405.12714}\ } (\bibinfo {year} {2024})}\BibitemShut {NoStop}%
\bibitem [{\citenamefont {Gonzalez-Conde}\ \emph {et~al.}(2024)\citenamefont {Gonzalez-Conde}, \citenamefont {Lewis}, \citenamefont {Bharadwaj},\ and\ \citenamefont {Sanz}}]{gonzalezconde2024}%
  \BibitemOpen
  \bibfield  {author} {\bibinfo {author} {\bibfnamefont {J.}~\bibnamefont {Gonzalez-Conde}}, \bibinfo {author} {\bibfnamefont {D.}~\bibnamefont {Lewis}}, \bibinfo {author} {\bibfnamefont {S.~S.}\ \bibnamefont {Bharadwaj}},\ and\ \bibinfo {author} {\bibfnamefont {M.}~\bibnamefont {Sanz}},\ }\href {https://arxiv.org/abs/2410.23057} {\bibinfo {title} {Quantum carleman linearisation efficiency in nonlinear fluid dynamics}} (\bibinfo {year} {2024}),\ \Eprint {https://arxiv.org/abs/2410.23057} {arXiv:2410.23057 [quant-ph]} \BibitemShut {NoStop}%
\bibitem [{\citenamefont {Sanavio}\ \emph {et~al.}(2024{\natexlab{a}})\citenamefont {Sanavio}, \citenamefont {Scatamacchia}, \citenamefont {de~Falco},\ and\ \citenamefont {Succi}}]{Sanavio2024}%
  \BibitemOpen
  \bibfield  {author} {\bibinfo {author} {\bibfnamefont {C.}~\bibnamefont {Sanavio}}, \bibinfo {author} {\bibfnamefont {R.}~\bibnamefont {Scatamacchia}}, \bibinfo {author} {\bibfnamefont {C.}~\bibnamefont {de~Falco}},\ and\ \bibinfo {author} {\bibfnamefont {S.}~\bibnamefont {Succi}},\ }\href {https://doi.org/10.1063/5.0204955} {\bibfield  {journal} {\bibinfo  {journal} {Physics of Fluids}\ }\textbf {\bibinfo {volume} {36}},\ \bibinfo {pages} {057143} (\bibinfo {year} {2024}{\natexlab{a}})}\BibitemShut {NoStop}%
\bibitem [{\citenamefont {Sanavio}\ \emph {et~al.}(2024{\natexlab{b}})\citenamefont {Sanavio}, \citenamefont {Mauri},\ and\ \citenamefont {Succi}}]{sanavio2024carleman}%
  \BibitemOpen
  \bibfield  {author} {\bibinfo {author} {\bibfnamefont {C.}~\bibnamefont {Sanavio}}, \bibinfo {author} {\bibfnamefont {E.}~\bibnamefont {Mauri}},\ and\ \bibinfo {author} {\bibfnamefont {S.}~\bibnamefont {Succi}},\ }\href {https://arxiv.org/abs/2406.01118} {\bibinfo {title} {Carleman-grad approach to the quantum simulation of fluids}} (\bibinfo {year} {2024}{\natexlab{b}}),\ \Eprint {https://arxiv.org/abs/2406.01118} {arXiv:2406.01118 [quant-ph]} \BibitemShut {NoStop}%
\bibitem [{\citenamefont {Paine}\ \emph {et~al.}(2023{\natexlab{a}})\citenamefont {Paine}, \citenamefont {Elfving},\ and\ \citenamefont {Kyriienko}}]{Paine2023}%
  \BibitemOpen
  \bibfield  {author} {\bibinfo {author} {\bibfnamefont {A.~E.}\ \bibnamefont {Paine}}, \bibinfo {author} {\bibfnamefont {V.~E.}\ \bibnamefont {Elfving}},\ and\ \bibinfo {author} {\bibfnamefont {O.}~\bibnamefont {Kyriienko}},\ }\href@noop {} {\bibinfo {title} {Physics-informed quantum machine learning: Solving nonlinear differential equations in latent spaces without costly grid evaluations}} (\bibinfo {year} {2023}{\natexlab{a}}),\ \Eprint {https://arxiv.org/abs/2308.01827} {arXiv:2308.01827 [quant-ph]} \BibitemShut {NoStop}%
\bibitem [{\citenamefont {Wu}\ \emph {et~al.}(2025)\citenamefont {Wu}, \citenamefont {Paine}, \citenamefont {Philip}, \citenamefont {Gentile},\ and\ \citenamefont {Kyriienko}}]{wu2025heff}%
  \BibitemOpen
  \bibfield  {author} {\bibinfo {author} {\bibfnamefont {H.-Y.}\ \bibnamefont {Wu}}, \bibinfo {author} {\bibfnamefont {A.~E.}\ \bibnamefont {Paine}}, \bibinfo {author} {\bibfnamefont {E.}~\bibnamefont {Philip}}, \bibinfo {author} {\bibfnamefont {A.~A.}\ \bibnamefont {Gentile}},\ and\ \bibinfo {author} {\bibfnamefont {O.}~\bibnamefont {Kyriienko}},\ }\href {https://arxiv.org/abs/2504.13174} {\bibinfo {title} {Quantum algorithm for solving nonlinear differential equations based on physics-informed effective hamiltonians}} (\bibinfo {year} {2025}),\ \Eprint {https://arxiv.org/abs/2504.13174} {arXiv:2504.13174 [quant-ph]} \BibitemShut {NoStop}%
\bibitem [{\citenamefont {Lubasch}\ \emph {et~al.}(2020)\citenamefont {Lubasch}, \citenamefont {Joo}, \citenamefont {Moinier}, \citenamefont {Kiffner},\ and\ \citenamefont {Jaksch}}]{lubasch2020variational}%
  \BibitemOpen
  \bibfield  {author} {\bibinfo {author} {\bibfnamefont {M.}~\bibnamefont {Lubasch}}, \bibinfo {author} {\bibfnamefont {J.}~\bibnamefont {Joo}}, \bibinfo {author} {\bibfnamefont {P.}~\bibnamefont {Moinier}}, \bibinfo {author} {\bibfnamefont {M.}~\bibnamefont {Kiffner}},\ and\ \bibinfo {author} {\bibfnamefont {D.}~\bibnamefont {Jaksch}},\ }\href {https://journals.aps.org/pra/abstract/10.1103/PhysRevA.101.010301} {\bibfield  {journal} {\bibinfo  {journal} {Physical Review A}\ }\textbf {\bibinfo {volume} {101}},\ \bibinfo {pages} {010301} (\bibinfo {year} {2020})}\BibitemShut {NoStop}%
\bibitem [{\citenamefont {Jaksch}\ \emph {et~al.}(2022)\citenamefont {Jaksch}, \citenamefont {Givi}, \citenamefont {Daley},\ and\ \citenamefont {Rung}}]{jaksch2022variational}%
  \BibitemOpen
  \bibfield  {author} {\bibinfo {author} {\bibfnamefont {D.}~\bibnamefont {Jaksch}}, \bibinfo {author} {\bibfnamefont {P.}~\bibnamefont {Givi}}, \bibinfo {author} {\bibfnamefont {A.~J.}\ \bibnamefont {Daley}},\ and\ \bibinfo {author} {\bibfnamefont {T.}~\bibnamefont {Rung}},\ }\href@noop {} {\bibinfo {title} {Variational quantum algorithms for computational fluid dynamics}} (\bibinfo {year} {2022}),\ \Eprint {https://arxiv.org/abs/2209.04915} {arXiv:2209.04915 [quant-ph]} \BibitemShut {NoStop}%
\bibitem [{\citenamefont {Pool}\ \emph {et~al.}(2024)\citenamefont {Pool}, \citenamefont {Somoza}, \citenamefont {Mc~Keever}, \citenamefont {Lubasch},\ and\ \citenamefont {Horstmann}}]{pool2024nonlinear}%
  \BibitemOpen
  \bibfield  {author} {\bibinfo {author} {\bibfnamefont {A.~J.}\ \bibnamefont {Pool}}, \bibinfo {author} {\bibfnamefont {A.~D.}\ \bibnamefont {Somoza}}, \bibinfo {author} {\bibfnamefont {C.}~\bibnamefont {Mc~Keever}}, \bibinfo {author} {\bibfnamefont {M.}~\bibnamefont {Lubasch}},\ and\ \bibinfo {author} {\bibfnamefont {B.}~\bibnamefont {Horstmann}},\ }\href@noop {} {\bibfield  {journal} {\bibinfo  {journal} {Physical Review Research}\ }\textbf {\bibinfo {volume} {6}},\ \bibinfo {pages} {033257} (\bibinfo {year} {2024})}\BibitemShut {NoStop}%
\bibitem [{\citenamefont {Biamonte}\ \emph {et~al.}(2017)\citenamefont {Biamonte}, \citenamefont {Wittek}, \citenamefont {Pancotti}, \citenamefont {Rebentrost}, \citenamefont {Wiebe},\ and\ \citenamefont {Lloyd}}]{Biamonte2017}%
  \BibitemOpen
  \bibfield  {author} {\bibinfo {author} {\bibfnamefont {J.}~\bibnamefont {Biamonte}}, \bibinfo {author} {\bibfnamefont {P.}~\bibnamefont {Wittek}}, \bibinfo {author} {\bibfnamefont {N.}~\bibnamefont {Pancotti}}, \bibinfo {author} {\bibfnamefont {P.}~\bibnamefont {Rebentrost}}, \bibinfo {author} {\bibfnamefont {N.}~\bibnamefont {Wiebe}},\ and\ \bibinfo {author} {\bibfnamefont {S.}~\bibnamefont {Lloyd}},\ }\href {https://doi.org/10.1038/nature23474} {\bibfield  {journal} {\bibinfo  {journal} {Nature}\ }\textbf {\bibinfo {volume} {549}},\ \bibinfo {pages} {195} (\bibinfo {year} {2017})}\BibitemShut {NoStop}%
\bibitem [{\citenamefont {Williams}\ \emph {et~al.}(2024{\natexlab{c}})\citenamefont {Williams}, \citenamefont {Scali}, \citenamefont {Gentile}, \citenamefont {Berger},\ and\ \citenamefont {Kyriienko}}]{Williams2024readout}%
  \BibitemOpen
  \bibfield  {author} {\bibinfo {author} {\bibfnamefont {C.~A.}\ \bibnamefont {Williams}}, \bibinfo {author} {\bibfnamefont {S.}~\bibnamefont {Scali}}, \bibinfo {author} {\bibfnamefont {A.~A.}\ \bibnamefont {Gentile}}, \bibinfo {author} {\bibfnamefont {D.}~\bibnamefont {Berger}},\ and\ \bibinfo {author} {\bibfnamefont {O.}~\bibnamefont {Kyriienko}},\ }\href {https://arxiv.org/abs/2411.14259} {\bibinfo {title} {Addressing the readout problem in quantum differential equation algorithms with quantum scientific machine learning}} (\bibinfo {year} {2024}{\natexlab{c}}),\ \Eprint {https://arxiv.org/abs/2411.14259} {arXiv:2411.14259 [quant-ph]} \BibitemShut {NoStop}%
\bibitem [{\citenamefont {Canivete~Cuissa}\ and\ \citenamefont {Steiner}(2022)}]{swirl_vortex_identification}%
  \BibitemOpen
  \bibfield  {author} {\bibinfo {author} {\bibfnamefont {J.~R.}\ \bibnamefont {Canivete~Cuissa}}\ and\ \bibinfo {author} {\bibfnamefont {O.}~\bibnamefont {Steiner}},\ }\href {https://doi.org/10.1051/0004-6361/202243740} {\bibfield  {journal} {\bibinfo  {journal} {Astronomy I\&; Astrophysics}\ }\textbf {\bibinfo {volume} {668}},\ \bibinfo {pages} {A118} (\bibinfo {year} {2022})}\BibitemShut {NoStop}%
\bibitem [{\citenamefont {Epps}(2017)}]{Epps2017}%
  \BibitemOpen
  \bibfield  {author} {\bibinfo {author} {\bibfnamefont {B.}~\bibnamefont {Epps}},\ }\bibinfo {title} {Review of vortex identification methods},\ in\ \href {https://doi.org/10.2514/6.2017-0989} {\emph {\bibinfo {booktitle} {55th AIAA Aerospace Sciences Meeting}}}\ (\bibinfo  {publisher} {American Institute of Aeronautics and Astronautics, Inc.},\ \bibinfo {year} {2017})\ \Eprint {https://arxiv.org/abs/https://arc.aiaa.org/doi/pdf/10.2514/6.2017-0989} {https://arc.aiaa.org/doi/pdf/10.2514/6.2017-0989} \BibitemShut {NoStop}%
\bibitem [{\citenamefont {Brunton}\ \emph {et~al.}(2020)\citenamefont {Brunton}, \citenamefont {Noack},\ and\ \citenamefont {Koumoutsakos}}]{Brunton2020rev}%
  \BibitemOpen
  \bibfield  {author} {\bibinfo {author} {\bibfnamefont {S.~L.}\ \bibnamefont {Brunton}}, \bibinfo {author} {\bibfnamefont {B.~R.}\ \bibnamefont {Noack}},\ and\ \bibinfo {author} {\bibfnamefont {P.}~\bibnamefont {Koumoutsakos}},\ }\href {https://doi.org/https://doi.org/10.1146/annurev-fluid-010719-060214} {\bibfield  {journal} {\bibinfo  {journal} {Annual Review of Fluid Mechanics}\ }\textbf {\bibinfo {volume} {52}},\ \bibinfo {pages} {477} (\bibinfo {year} {2020})}\BibitemShut {NoStop}%
\bibitem [{\citenamefont {Duraisamy}\ \emph {et~al.}(2019)\citenamefont {Duraisamy}, \citenamefont {Iaccarino},\ and\ \citenamefont {Xiao}}]{Duraisamy_annurev}%
  \BibitemOpen
  \bibfield  {author} {\bibinfo {author} {\bibfnamefont {K.}~\bibnamefont {Duraisamy}}, \bibinfo {author} {\bibfnamefont {G.}~\bibnamefont {Iaccarino}},\ and\ \bibinfo {author} {\bibfnamefont {H.}~\bibnamefont {Xiao}},\ }\href {https://doi.org/https://doi.org/10.1146/annurev-fluid-010518-040547} {\bibfield  {journal} {\bibinfo  {journal} {Annual Review of Fluid Mechanics}\ }\textbf {\bibinfo {volume} {51}},\ \bibinfo {pages} {357} (\bibinfo {year} {2019})}\BibitemShut {NoStop}%
\bibitem [{\citenamefont {Ströfer}\ \emph {et~al.}(2019)\citenamefont {Ströfer}, \citenamefont {Wu}, \citenamefont {Xiao},\ and\ \citenamefont {Paterson}}]{cnn_vortex_identification}%
  \BibitemOpen
  \bibfield  {author} {\bibinfo {author} {\bibfnamefont {C.}~\bibnamefont {Ströfer}, \bibfnamefont {Michelén}}, \bibinfo {author} {\bibfnamefont {J.-L.}\ \bibnamefont {Wu}}, \bibinfo {author} {\bibfnamefont {H.}~\bibnamefont {Xiao}},\ and\ \bibinfo {author} {\bibfnamefont {E.}~\bibnamefont {Paterson}},\ }\href {https://doi.org/https://doi.org/10.4208/cicp.OA-2018-0035} {\bibfield  {journal} {\bibinfo  {journal} {Communications in Computational Physics}\ }\textbf {\bibinfo {volume} {25}},\ \bibinfo {pages} {625} (\bibinfo {year} {2019})}\BibitemShut {NoStop}%
\bibitem [{\citenamefont {Wang}\ \emph {et~al.}(2021)\citenamefont {Wang}, \citenamefont {Liang}, \citenamefont {Yang}, \citenamefont {Zhao},\ and\ \citenamefont {Wang}}]{vortex_seg_net}%
  \BibitemOpen
  \bibfield  {author} {\bibinfo {author} {\bibfnamefont {Y.}~\bibnamefont {Wang}}, \bibinfo {author} {\bibfnamefont {D.}~\bibnamefont {Liang}}, \bibinfo {author} {\bibfnamefont {Z.}~\bibnamefont {Yang}}, \bibinfo {author} {\bibfnamefont {D.}~\bibnamefont {Zhao}},\ and\ \bibinfo {author} {\bibfnamefont {F.}~\bibnamefont {Wang}},\ }\href {https://doi.org/10.1007/s00371-020-01797-6} {\bibfield  {journal} {\bibinfo  {journal} {The Visual Computer}\ }\textbf {\bibinfo {volume} {37}} (\bibinfo {year} {2021})}\BibitemShut {NoStop}%
\bibitem [{\citenamefont {Deng}\ \emph {et~al.}(2022)\citenamefont {Deng}, \citenamefont {Bao}, \citenamefont {Wang}, \citenamefont {Yang}, \citenamefont {Zhao}, \citenamefont {Wang}, \citenamefont {Bi},\ and\ \citenamefont {Guo}}]{vortex_u_net}%
  \BibitemOpen
  \bibfield  {author} {\bibinfo {author} {\bibfnamefont {L.}~\bibnamefont {Deng}}, \bibinfo {author} {\bibfnamefont {W.}~\bibnamefont {Bao}}, \bibinfo {author} {\bibfnamefont {Y.}~\bibnamefont {Wang}}, \bibinfo {author} {\bibfnamefont {Z.}~\bibnamefont {Yang}}, \bibinfo {author} {\bibfnamefont {D.}~\bibnamefont {Zhao}}, \bibinfo {author} {\bibfnamefont {F.}~\bibnamefont {Wang}}, \bibinfo {author} {\bibfnamefont {C.}~\bibnamefont {Bi}},\ and\ \bibinfo {author} {\bibfnamefont {Y.}~\bibnamefont {Guo}},\ }\href {https://doi.org/https://doi.org/10.1016/j.asoc.2021.108229} {\bibfield  {journal} {\bibinfo  {journal} {Applied Soft Computing}\ }\textbf {\bibinfo {volume} {115}},\ \bibinfo {pages} {108229} (\bibinfo {year} {2022})}\BibitemShut {NoStop}%
\bibitem [{\citenamefont {Liang}\ \emph {et~al.}(2018)\citenamefont {Liang}, \citenamefont {Wang}, \citenamefont {Liu}, \citenamefont {Wang}, \citenamefont {Li},\ and\ \citenamefont {Liu}}]{vortex_cnn}%
  \BibitemOpen
  \bibfield  {author} {\bibinfo {author} {\bibfnamefont {D.}~\bibnamefont {Liang}}, \bibinfo {author} {\bibfnamefont {Y.}~\bibnamefont {Wang}}, \bibinfo {author} {\bibfnamefont {Y.}~\bibnamefont {Liu}}, \bibinfo {author} {\bibfnamefont {F.}~\bibnamefont {Wang}}, \bibinfo {author} {\bibfnamefont {S.}~\bibnamefont {Li}},\ and\ \bibinfo {author} {\bibfnamefont {J.}~\bibnamefont {Liu}},\ }\href {https://doi.org/10.1007/s12650-018-0523-1} {\bibfield  {journal} {\bibinfo  {journal} {Journal of Visualization}\ }\textbf {\bibinfo {volume} {22}} (\bibinfo {year} {2018})}\BibitemShut {NoStop}%
\bibitem [{\citenamefont {Wang}\ \emph {et~al.}(2020)\citenamefont {Wang}, \citenamefont {Guo}, \citenamefont {Wang}, \citenamefont {Deng}, \citenamefont {Wang},\ and\ \citenamefont {Li}}]{vortex_elm_net}%
  \BibitemOpen
  \bibfield  {author} {\bibinfo {author} {\bibfnamefont {J.}~\bibnamefont {Wang}}, \bibinfo {author} {\bibfnamefont {L.}~\bibnamefont {Guo}}, \bibinfo {author} {\bibfnamefont {Y.}~\bibnamefont {Wang}}, \bibinfo {author} {\bibfnamefont {L.}~\bibnamefont {Deng}}, \bibinfo {author} {\bibfnamefont {F.}~\bibnamefont {Wang}},\ and\ \bibinfo {author} {\bibfnamefont {T.}~\bibnamefont {Li}},\ }\href {https://doi.org/https://doi.org/10.1155/2020/8865001} {\bibfield  {journal} {\bibinfo  {journal} {International Journal of Aerospace Engineering}\ }\textbf {\bibinfo {volume} {2020}},\ \bibinfo {pages} {8865001} (\bibinfo {year} {2020})}\BibitemShut {NoStop}%
\bibitem [{\citenamefont {Huang}\ \emph {et~al.}(2022)\citenamefont {Huang}, \citenamefont {Broughton}, \citenamefont {Cotler}, \citenamefont {Chen}, \citenamefont {Li}, \citenamefont {Mohseni}, \citenamefont {Neven}, \citenamefont {Babbush}, \citenamefont {Kueng}, \citenamefont {Preskill},\ and\ \citenamefont {McClean}}]{Huang2022}%
  \BibitemOpen
  \bibfield  {author} {\bibinfo {author} {\bibfnamefont {H.-Y.}\ \bibnamefont {Huang}}, \bibinfo {author} {\bibfnamefont {M.}~\bibnamefont {Broughton}}, \bibinfo {author} {\bibfnamefont {J.}~\bibnamefont {Cotler}}, \bibinfo {author} {\bibfnamefont {S.}~\bibnamefont {Chen}}, \bibinfo {author} {\bibfnamefont {J.}~\bibnamefont {Li}}, \bibinfo {author} {\bibfnamefont {M.}~\bibnamefont {Mohseni}}, \bibinfo {author} {\bibfnamefont {H.}~\bibnamefont {Neven}}, \bibinfo {author} {\bibfnamefont {R.}~\bibnamefont {Babbush}}, \bibinfo {author} {\bibfnamefont {R.}~\bibnamefont {Kueng}}, \bibinfo {author} {\bibfnamefont {J.}~\bibnamefont {Preskill}},\ and\ \bibinfo {author} {\bibfnamefont {J.~R.}\ \bibnamefont {McClean}},\ }\href {https://doi.org/10.1126/science.abn7293} {\bibfield  {journal} {\bibinfo  {journal} {Science}\ }\textbf {\bibinfo {volume} {376}},\ \bibinfo {pages} {1182} (\bibinfo {year} {2022})}\BibitemShut {NoStop}%
\bibitem [{\citenamefont {Schatzki}\ \emph {et~al.}(2021)\citenamefont {Schatzki}, \citenamefont {Arrasmith}, \citenamefont {Coles},\ and\ \citenamefont {Cerezo}}]{schatzki2021entangled}%
  \BibitemOpen
  \bibfield  {author} {\bibinfo {author} {\bibfnamefont {L.}~\bibnamefont {Schatzki}}, \bibinfo {author} {\bibfnamefont {A.}~\bibnamefont {Arrasmith}}, \bibinfo {author} {\bibfnamefont {P.~J.}\ \bibnamefont {Coles}},\ and\ \bibinfo {author} {\bibfnamefont {M.}~\bibnamefont {Cerezo}},\ }\href@noop {} {\bibinfo {title} {Entangled datasets for quantum machine learning}} (\bibinfo {year} {2021}),\ \Eprint {https://arxiv.org/abs/2109.03400} {arXiv:2109.03400 [quant-ph]} \BibitemShut {NoStop}%
\bibitem [{\citenamefont {Caro}\ \emph {et~al.}(2022)\citenamefont {Caro}, \citenamefont {Huang}, \citenamefont {Cerezo}, \citenamefont {Sharma}, \citenamefont {Sornborger}, \citenamefont {Cincio},\ and\ \citenamefont {Coles}}]{Caro2022}%
  \BibitemOpen
  \bibfield  {author} {\bibinfo {author} {\bibfnamefont {M.~C.}\ \bibnamefont {Caro}}, \bibinfo {author} {\bibfnamefont {H.-Y.}\ \bibnamefont {Huang}}, \bibinfo {author} {\bibfnamefont {M.}~\bibnamefont {Cerezo}}, \bibinfo {author} {\bibfnamefont {K.}~\bibnamefont {Sharma}}, \bibinfo {author} {\bibfnamefont {A.}~\bibnamefont {Sornborger}}, \bibinfo {author} {\bibfnamefont {L.}~\bibnamefont {Cincio}},\ and\ \bibinfo {author} {\bibfnamefont {P.~J.}\ \bibnamefont {Coles}},\ }\href {https://doi.org/10.1038/s41467-022-32550-3} {\bibfield  {journal} {\bibinfo  {journal} {Nature Communications}\ }\textbf {\bibinfo {volume} {13}},\ \bibinfo {pages} {4919} (\bibinfo {year} {2022})}\BibitemShut {NoStop}%
\bibitem [{\citenamefont {Anschuetz}\ \emph {et~al.}(2023)\citenamefont {Anschuetz}, \citenamefont {Hu}, \citenamefont {Huang},\ and\ \citenamefont {Gao}}]{Anschuetz2023}%
  \BibitemOpen
  \bibfield  {author} {\bibinfo {author} {\bibfnamefont {E.~R.}\ \bibnamefont {Anschuetz}}, \bibinfo {author} {\bibfnamefont {H.-Y.}\ \bibnamefont {Hu}}, \bibinfo {author} {\bibfnamefont {J.-L.}\ \bibnamefont {Huang}},\ and\ \bibinfo {author} {\bibfnamefont {X.}~\bibnamefont {Gao}},\ }\href {https://doi.org/10.1103/PRXQuantum.4.020338} {\bibfield  {journal} {\bibinfo  {journal} {PRX Quantum}\ }\textbf {\bibinfo {volume} {4}},\ \bibinfo {pages} {020338} (\bibinfo {year} {2023})}\BibitemShut {NoStop}%
\bibitem [{\citenamefont {Umeano}\ \emph {et~al.}(2024{\natexlab{a}})\citenamefont {Umeano}, \citenamefont {Elfving},\ and\ \citenamefont {Kyriienko}}]{umeano2024geometric}%
  \BibitemOpen
  \bibfield  {author} {\bibinfo {author} {\bibfnamefont {C.}~\bibnamefont {Umeano}}, \bibinfo {author} {\bibfnamefont {V.~E.}\ \bibnamefont {Elfving}},\ and\ \bibinfo {author} {\bibfnamefont {O.}~\bibnamefont {Kyriienko}},\ }\href {https://arxiv.org/abs/2402.03871} {\bibinfo {title} {Geometric quantum machine learning of bqp$^a$ protocols and latent graph classifiers}} (\bibinfo {year} {2024}{\natexlab{a}}),\ \Eprint {https://arxiv.org/abs/2402.03871} {arXiv:2402.03871 [quant-ph]} \BibitemShut {NoStop}%
\bibitem [{\citenamefont {Umeano}\ \emph {et~al.}(2024{\natexlab{b}})\citenamefont {Umeano}, \citenamefont {Scali},\ and\ \citenamefont {Kyriienko}}]{umeano2024forrelation}%
  \BibitemOpen
  \bibfield  {author} {\bibinfo {author} {\bibfnamefont {C.}~\bibnamefont {Umeano}}, \bibinfo {author} {\bibfnamefont {S.}~\bibnamefont {Scali}},\ and\ \bibinfo {author} {\bibfnamefont {O.}~\bibnamefont {Kyriienko}},\ }\href {https://arxiv.org/abs/2409.01496} {\bibinfo {title} {Can geometric quantum machine learning lead to advantage in barcode classification?}} (\bibinfo {year} {2024}{\natexlab{b}}),\ \Eprint {https://arxiv.org/abs/2409.01496} {arXiv:2409.01496 [quant-ph]} \BibitemShut {NoStop}%
\bibitem [{\citenamefont {Lewis}\ \emph {et~al.}(2025)\citenamefont {Lewis}, \citenamefont {Gilboa},\ and\ \citenamefont {McClean}}]{lewis2025functions}%
  \BibitemOpen
  \bibfield  {author} {\bibinfo {author} {\bibfnamefont {L.}~\bibnamefont {Lewis}}, \bibinfo {author} {\bibfnamefont {D.}~\bibnamefont {Gilboa}},\ and\ \bibinfo {author} {\bibfnamefont {J.~R.}\ \bibnamefont {McClean}},\ }\href {https://arxiv.org/abs/2503.20879} {\bibinfo {title} {Quantum advantage for learning shallow neural networks with natural data distributions}} (\bibinfo {year} {2025}),\ \Eprint {https://arxiv.org/abs/2503.20879} {arXiv:2503.20879 [quant-ph]} \BibitemShut {NoStop}%
\bibitem [{\citenamefont {Morohoshi}\ \emph {et~al.}(2025)\citenamefont {Morohoshi}, \citenamefont {Nakayama}, \citenamefont {Manabe},\ and\ \citenamefont {Mitarai}}]{morohoshi2025hamiltonians}%
  \BibitemOpen
  \bibfield  {author} {\bibinfo {author} {\bibfnamefont {Y.}~\bibnamefont {Morohoshi}}, \bibinfo {author} {\bibfnamefont {A.}~\bibnamefont {Nakayama}}, \bibinfo {author} {\bibfnamefont {H.}~\bibnamefont {Manabe}},\ and\ \bibinfo {author} {\bibfnamefont {K.}~\bibnamefont {Mitarai}},\ }\href {https://arxiv.org/abs/2504.16370} {\bibinfo {title} {Learning functions of hamiltonians with hamiltonian fourier features}} (\bibinfo {year} {2025}),\ \Eprint {https://arxiv.org/abs/2504.16370} {arXiv:2504.16370 [quant-ph]} \BibitemShut {NoStop}%
\bibitem [{\citenamefont {Barthe}\ \emph {et~al.}(2025)\citenamefont {Barthe}, \citenamefont {Rad}, \citenamefont {Grossi},\ and\ \citenamefont {Dunjko}}]{barthe2025dynamics}%
  \BibitemOpen
  \bibfield  {author} {\bibinfo {author} {\bibfnamefont {A.}~\bibnamefont {Barthe}}, \bibinfo {author} {\bibfnamefont {M.~Y.}\ \bibnamefont {Rad}}, \bibinfo {author} {\bibfnamefont {M.}~\bibnamefont {Grossi}},\ and\ \bibinfo {author} {\bibfnamefont {V.}~\bibnamefont {Dunjko}},\ }\href {https://arxiv.org/abs/2506.17089} {\bibinfo {title} {Quantum advantage in learning quantum dynamics via fourier coefficient extraction}} (\bibinfo {year} {2025}),\ \Eprint {https://arxiv.org/abs/2506.17089} {arXiv:2506.17089 [quant-ph]} \BibitemShut {NoStop}%
\bibitem [{\citenamefont {Kyriienko}\ \emph {et~al.}(2021)\citenamefont {Kyriienko}, \citenamefont {Paine},\ and\ \citenamefont {Elfving}}]{kyriienko2021solving}%
  \BibitemOpen
  \bibfield  {author} {\bibinfo {author} {\bibfnamefont {O.}~\bibnamefont {Kyriienko}}, \bibinfo {author} {\bibfnamefont {A.~E.}\ \bibnamefont {Paine}},\ and\ \bibinfo {author} {\bibfnamefont {V.~E.}\ \bibnamefont {Elfving}},\ }\href {https://doi.org/10.1103/PhysRevA.103.052416} {\bibfield  {journal} {\bibinfo  {journal} {Phys. Rev. A}\ }\textbf {\bibinfo {volume} {103}},\ \bibinfo {pages} {052416} (\bibinfo {year} {2021})}\BibitemShut {NoStop}%
\bibitem [{\citenamefont {Paine}\ \emph {et~al.}(2023{\natexlab{b}})\citenamefont {Paine}, \citenamefont {Elfving},\ and\ \citenamefont {Kyriienko}}]{Paine2021}%
  \BibitemOpen
  \bibfield  {author} {\bibinfo {author} {\bibfnamefont {A.~E.}\ \bibnamefont {Paine}}, \bibinfo {author} {\bibfnamefont {V.~E.}\ \bibnamefont {Elfving}},\ and\ \bibinfo {author} {\bibfnamefont {O.}~\bibnamefont {Kyriienko}},\ }\href {https://doi.org/https://doi.org/10.1002/qute.202300065} {\bibfield  {journal} {\bibinfo  {journal} {Advanced Quantum Technologies}\ }\textbf {\bibinfo {volume} {6}},\ \bibinfo {pages} {2300065} (\bibinfo {year} {2023}{\natexlab{b}})}\BibitemShut {NoStop}%
\bibitem [{\citenamefont {Paine}\ \emph {et~al.}(2023{\natexlab{c}})\citenamefont {Paine}, \citenamefont {Elfving},\ and\ \citenamefont {Kyriienko}}]{paine2023quantum}%
  \BibitemOpen
  \bibfield  {author} {\bibinfo {author} {\bibfnamefont {A.~E.}\ \bibnamefont {Paine}}, \bibinfo {author} {\bibfnamefont {V.~E.}\ \bibnamefont {Elfving}},\ and\ \bibinfo {author} {\bibfnamefont {O.}~\bibnamefont {Kyriienko}},\ }\href {https://doi.org/10.1103/PhysRevA.107.032428} {\bibfield  {journal} {\bibinfo  {journal} {Physical Review A}\ }\textbf {\bibinfo {volume} {107}},\ \bibinfo {pages} {032428} (\bibinfo {year} {2023}{\natexlab{c}})}\BibitemShut {NoStop}%
\bibitem [{\citenamefont {Markidis}(2022)}]{Markidis2022qpinns}%
  \BibitemOpen
  \bibfield  {author} {\bibinfo {author} {\bibfnamefont {S.}~\bibnamefont {Markidis}},\ }\href {https://www.frontiersin.org/articles/10.3389/fams.2022.1036711} {\bibfield  {journal} {\bibinfo  {journal} {Frontiers in Applied Mathematics and Statistics}\ }\textbf {\bibinfo {volume} {8}} (\bibinfo {year} {2022})}\BibitemShut {NoStop}%
\bibitem [{\citenamefont {Heim}\ \emph {et~al.}(2021)\citenamefont {Heim}, \citenamefont {Ghosh}, \citenamefont {Kyriienko},\ and\ \citenamefont {Elfving}}]{heim2021quantum}%
  \BibitemOpen
  \bibfield  {author} {\bibinfo {author} {\bibfnamefont {N.}~\bibnamefont {Heim}}, \bibinfo {author} {\bibfnamefont {A.}~\bibnamefont {Ghosh}}, \bibinfo {author} {\bibfnamefont {O.}~\bibnamefont {Kyriienko}},\ and\ \bibinfo {author} {\bibfnamefont {V.~E.}\ \bibnamefont {Elfving}},\ }\href@noop {} {\bibinfo {title} {Quantum model-discovery}} (\bibinfo {year} {2021}),\ \Eprint {https://arxiv.org/abs/2111.06376} {arXiv:2111.06376 [quant-ph]} \BibitemShut {NoStop}%
\bibitem [{\citenamefont {Kasture}\ \emph {et~al.}(2023)\citenamefont {Kasture}, \citenamefont {Kyriienko},\ and\ \citenamefont {Elfving}}]{kasture2022protocols}%
  \BibitemOpen
  \bibfield  {author} {\bibinfo {author} {\bibfnamefont {S.}~\bibnamefont {Kasture}}, \bibinfo {author} {\bibfnamefont {O.}~\bibnamefont {Kyriienko}},\ and\ \bibinfo {author} {\bibfnamefont {V.~E.}\ \bibnamefont {Elfving}},\ }\href {https://doi.org/10.1103/PhysRevA.108.042406} {\bibfield  {journal} {\bibinfo  {journal} {Phys. Rev. A}\ }\textbf {\bibinfo {volume} {108}},\ \bibinfo {pages} {042406} (\bibinfo {year} {2023})}\BibitemShut {NoStop}%
\bibitem [{\citenamefont {Setty}\ \emph {et~al.}(2023)\citenamefont {Setty}, \citenamefont {Abdusalamov},\ and\ \citenamefont {Itskov}}]{pinn_qml}%
  \BibitemOpen
  \bibfield  {author} {\bibinfo {author} {\bibfnamefont {A.}~\bibnamefont {Setty}}, \bibinfo {author} {\bibfnamefont {R.}~\bibnamefont {Abdusalamov}},\ and\ \bibinfo {author} {\bibfnamefont {M.}~\bibnamefont {Itskov}},\ }\href {https://arxiv.org/abs/2312.09215} {\bibinfo {title} {Physics-informed quantum machine learning for solving partial differential equations}} (\bibinfo {year} {2023}),\ \Eprint {https://arxiv.org/abs/2312.09215} {arXiv:2312.09215 [quant-ph]} \BibitemShut {NoStop}%
\bibitem [{\citenamefont {Jaderberg}\ \emph {et~al.}(2024{\natexlab{a}})\citenamefont {Jaderberg}, \citenamefont {Gentile}, \citenamefont {Berrada}, \citenamefont {Shishenina},\ and\ \citenamefont {Elfving}}]{Jaderberg2024}%
  \BibitemOpen
  \bibfield  {author} {\bibinfo {author} {\bibfnamefont {B.}~\bibnamefont {Jaderberg}}, \bibinfo {author} {\bibfnamefont {A.~A.}\ \bibnamefont {Gentile}}, \bibinfo {author} {\bibfnamefont {Y.~A.}\ \bibnamefont {Berrada}}, \bibinfo {author} {\bibfnamefont {E.}~\bibnamefont {Shishenina}},\ and\ \bibinfo {author} {\bibfnamefont {V.~E.}\ \bibnamefont {Elfving}},\ }\href {https://doi.org/10.1103/PhysRevA.109.042421} {\bibfield  {journal} {\bibinfo  {journal} {Phys. Rev. A}\ }\textbf {\bibinfo {volume} {109}},\ \bibinfo {pages} {042421} (\bibinfo {year} {2024}{\natexlab{a}})}\BibitemShut {NoStop}%
\bibitem [{\citenamefont {Jaderberg}\ \emph {et~al.}(2024{\natexlab{b}})\citenamefont {Jaderberg}, \citenamefont {Gentile}, \citenamefont {Ghosh}, \citenamefont {Elfving}, \citenamefont {Jones}, \citenamefont {Vodola}, \citenamefont {Manobianco},\ and\ \citenamefont {Weiss}}]{Jaderberg2024weather}%
  \BibitemOpen
  \bibfield  {author} {\bibinfo {author} {\bibfnamefont {B.}~\bibnamefont {Jaderberg}}, \bibinfo {author} {\bibfnamefont {A.~A.}\ \bibnamefont {Gentile}}, \bibinfo {author} {\bibfnamefont {A.}~\bibnamefont {Ghosh}}, \bibinfo {author} {\bibfnamefont {V.~E.}\ \bibnamefont {Elfving}}, \bibinfo {author} {\bibfnamefont {C.}~\bibnamefont {Jones}}, \bibinfo {author} {\bibfnamefont {D.}~\bibnamefont {Vodola}}, \bibinfo {author} {\bibfnamefont {J.}~\bibnamefont {Manobianco}},\ and\ \bibinfo {author} {\bibfnamefont {H.}~\bibnamefont {Weiss}},\ }\href {https://doi.org/10.1103/PhysRevA.110.052423} {\bibfield  {journal} {\bibinfo  {journal} {Phys. Rev. A}\ }\textbf {\bibinfo {volume} {110}},\ \bibinfo {pages} {052423} (\bibinfo {year} {2024}{\natexlab{b}})}\BibitemShut {NoStop}%
\bibitem [{\citenamefont {de~Lejarza}\ \emph {et~al.}(2025)\citenamefont {de~Lejarza}, \citenamefont {Wu}, \citenamefont {Kyriienko}, \citenamefont {Rodrigo},\ and\ \citenamefont {Grossi}}]{delejarza2025QCPM}%
  \BibitemOpen
  \bibfield  {author} {\bibinfo {author} {\bibfnamefont {J.~J.~M.}\ \bibnamefont {de~Lejarza}}, \bibinfo {author} {\bibfnamefont {H.-Y.}\ \bibnamefont {Wu}}, \bibinfo {author} {\bibfnamefont {O.}~\bibnamefont {Kyriienko}}, \bibinfo {author} {\bibfnamefont {G.}~\bibnamefont {Rodrigo}},\ and\ \bibinfo {author} {\bibfnamefont {M.}~\bibnamefont {Grossi}},\ }\href {https://arxiv.org/abs/2503.16073} {\bibinfo {title} {Quantum chebyshev probabilistic models for fragmentation functions}} (\bibinfo {year} {2025}),\ \Eprint {https://arxiv.org/abs/2503.16073} {arXiv:2503.16073 [quant-ph]} \BibitemShut {NoStop}%
\bibitem [{\citenamefont {Cong}\ \emph {et~al.}(2019)\citenamefont {Cong}, \citenamefont {Choi},\ and\ \citenamefont {Lukin}}]{Cong2019}%
  \BibitemOpen
  \bibfield  {author} {\bibinfo {author} {\bibfnamefont {I.}~\bibnamefont {Cong}}, \bibinfo {author} {\bibfnamefont {S.}~\bibnamefont {Choi}},\ and\ \bibinfo {author} {\bibfnamefont {M.~D.}\ \bibnamefont {Lukin}},\ }\href {https://doi.org/10.1038/s41567-019-0648-8} {\bibfield  {journal} {\bibinfo  {journal} {Nature Physics}\ }\textbf {\bibinfo {volume} {15}},\ \bibinfo {pages} {1273–1278} (\bibinfo {year} {2019})}\BibitemShut {NoStop}%
\bibitem [{\citenamefont {Pesah}\ \emph {et~al.}(2021)\citenamefont {Pesah}, \citenamefont {Cerezo}, \citenamefont {Wang}, \citenamefont {Volkoff}, \citenamefont {Sornborger},\ and\ \citenamefont {Coles}}]{pesah2021absence}%
  \BibitemOpen
  \bibfield  {author} {\bibinfo {author} {\bibfnamefont {A.}~\bibnamefont {Pesah}}, \bibinfo {author} {\bibfnamefont {M.}~\bibnamefont {Cerezo}}, \bibinfo {author} {\bibfnamefont {S.}~\bibnamefont {Wang}}, \bibinfo {author} {\bibfnamefont {T.}~\bibnamefont {Volkoff}}, \bibinfo {author} {\bibfnamefont {A.~T.}\ \bibnamefont {Sornborger}},\ and\ \bibinfo {author} {\bibfnamefont {P.~J.}\ \bibnamefont {Coles}},\ }\href@noop {} {\bibfield  {journal} {\bibinfo  {journal} {Physical Review X}\ }\textbf {\bibinfo {volume} {11}},\ \bibinfo {pages} {041011} (\bibinfo {year} {2021})}\BibitemShut {NoStop}%
\bibitem [{\citenamefont {Hur}\ \emph {et~al.}(2022)\citenamefont {Hur}, \citenamefont {Kim},\ and\ \citenamefont {Park}}]{Hur2022}%
  \BibitemOpen
  \bibfield  {author} {\bibinfo {author} {\bibfnamefont {T.}~\bibnamefont {Hur}}, \bibinfo {author} {\bibfnamefont {L.}~\bibnamefont {Kim}},\ and\ \bibinfo {author} {\bibfnamefont {D.~K.}\ \bibnamefont {Park}},\ }\bibfield  {journal} {\bibinfo  {journal} {Quantum Machine Intelligence}\ }\textbf {\bibinfo {volume} {4}},\ \href {https://doi.org/10.1007/s42484-021-00061-x} {10.1007/s42484-021-00061-x} (\bibinfo {year} {2022})\BibitemShut {NoStop}%
\bibitem [{\citenamefont {Chen}\ \emph {et~al.}(2022)\citenamefont {Chen}, \citenamefont {Wei}, \citenamefont {Zhang}, \citenamefont {Yu},\ and\ \citenamefont {Yoo}}]{SChen2022}%
  \BibitemOpen
  \bibfield  {author} {\bibinfo {author} {\bibfnamefont {S.~Y.-C.}\ \bibnamefont {Chen}}, \bibinfo {author} {\bibfnamefont {T.-C.}\ \bibnamefont {Wei}}, \bibinfo {author} {\bibfnamefont {C.}~\bibnamefont {Zhang}}, \bibinfo {author} {\bibfnamefont {H.}~\bibnamefont {Yu}},\ and\ \bibinfo {author} {\bibfnamefont {S.}~\bibnamefont {Yoo}},\ }\href {https://doi.org/10.1103/PhysRevResearch.4.013231} {\bibfield  {journal} {\bibinfo  {journal} {Phys. Rev. Res.}\ }\textbf {\bibinfo {volume} {4}},\ \bibinfo {pages} {013231} (\bibinfo {year} {2022})}\BibitemShut {NoStop}%
\bibitem [{\citenamefont {Herrmann}\ \emph {et~al.}(2022)\citenamefont {Herrmann}, \citenamefont {Llima}, \citenamefont {Remm}, \citenamefont {Zapletal}, \citenamefont {McMahon}, \citenamefont {Scarato}, \citenamefont {Swiadek}, \citenamefont {Andersen}, \citenamefont {Hellings}, \citenamefont {Krinner} \emph {et~al.}}]{herrmann2022realizing}%
  \BibitemOpen
  \bibfield  {author} {\bibinfo {author} {\bibfnamefont {J.}~\bibnamefont {Herrmann}}, \bibinfo {author} {\bibfnamefont {S.~M.}\ \bibnamefont {Llima}}, \bibinfo {author} {\bibfnamefont {A.}~\bibnamefont {Remm}}, \bibinfo {author} {\bibfnamefont {P.}~\bibnamefont {Zapletal}}, \bibinfo {author} {\bibfnamefont {N.~A.}\ \bibnamefont {McMahon}}, \bibinfo {author} {\bibfnamefont {C.}~\bibnamefont {Scarato}}, \bibinfo {author} {\bibfnamefont {F.}~\bibnamefont {Swiadek}}, \bibinfo {author} {\bibfnamefont {C.~K.}\ \bibnamefont {Andersen}}, \bibinfo {author} {\bibfnamefont {C.}~\bibnamefont {Hellings}}, \bibinfo {author} {\bibfnamefont {S.}~\bibnamefont {Krinner}}, \emph {et~al.},\ }\href@noop {} {\bibfield  {journal} {\bibinfo  {journal} {Nature communications}\ }\textbf {\bibinfo {volume} {13}},\ \bibinfo {pages} {4144} (\bibinfo {year} {2022})}\BibitemShut {NoStop}%
\bibitem [{\citenamefont {Zapletal}\ \emph {et~al.}(2024)\citenamefont {Zapletal}, \citenamefont {McMahon},\ and\ \citenamefont {Hartmann}}]{Zapletal2024}%
  \BibitemOpen
  \bibfield  {author} {\bibinfo {author} {\bibfnamefont {P.}~\bibnamefont {Zapletal}}, \bibinfo {author} {\bibfnamefont {N.~A.}\ \bibnamefont {McMahon}},\ and\ \bibinfo {author} {\bibfnamefont {M.~J.}\ \bibnamefont {Hartmann}},\ }\href {https://doi.org/10.1103/PhysRevResearch.6.033111} {\bibfield  {journal} {\bibinfo  {journal} {Phys. Rev. Res.}\ }\textbf {\bibinfo {volume} {6}},\ \bibinfo {pages} {033111} (\bibinfo {year} {2024})}\BibitemShut {NoStop}%
\bibitem [{\citenamefont {Umeano}\ \emph {et~al.}(2024{\natexlab{c}})\citenamefont {Umeano}, \citenamefont {Paine}, \citenamefont {Elfving},\ and\ \citenamefont {Kyriienko}}]{Umeano2024QCNN}%
  \BibitemOpen
  \bibfield  {author} {\bibinfo {author} {\bibfnamefont {C.}~\bibnamefont {Umeano}}, \bibinfo {author} {\bibfnamefont {A.~E.}\ \bibnamefont {Paine}}, \bibinfo {author} {\bibfnamefont {V.~E.}\ \bibnamefont {Elfving}},\ and\ \bibinfo {author} {\bibfnamefont {O.}~\bibnamefont {Kyriienko}},\ }\href {https://doi.org/https://doi.org/10.1002/qute.202400325} {\bibfield  {journal} {\bibinfo  {journal} {Advanced Quantum Technologies}\ }\textbf {\bibinfo {volume} {n/a}},\ \bibinfo {pages} {2400325} (\bibinfo {year} {2024}{\natexlab{c}})},\ \Eprint {https://arxiv.org/abs/https://advanced.onlinelibrary.wiley.com/doi/pdf/10.1002/qute.202400325} {https://advanced.onlinelibrary.wiley.com/doi/pdf/10.1002/qute.202400325} \BibitemShut {NoStop}%
\bibitem [{\citenamefont {Gil-Fuster}\ \emph {et~al.}(2024)\citenamefont {Gil-Fuster}, \citenamefont {Eisert},\ and\ \citenamefont {Bravo-Prieto}}]{Gil-Fuster2024}%
  \BibitemOpen
  \bibfield  {author} {\bibinfo {author} {\bibfnamefont {E.}~\bibnamefont {Gil-Fuster}}, \bibinfo {author} {\bibfnamefont {J.}~\bibnamefont {Eisert}},\ and\ \bibinfo {author} {\bibfnamefont {C.}~\bibnamefont {Bravo-Prieto}},\ }\href {https://doi.org/10.1038/s41467-024-45882-z} {\bibfield  {journal} {\bibinfo  {journal} {Nature Communications}\ }\textbf {\bibinfo {volume} {15}},\ \bibinfo {pages} {2277} (\bibinfo {year} {2024})}\BibitemShut {NoStop}%
\bibitem [{\citenamefont {Song}\ \emph {et~al.}(2024)\citenamefont {Song}, \citenamefont {Li}, \citenamefont {Wu}, \citenamefont {Qin}, \citenamefont {Wen},\ and\ \citenamefont {Gao}}]{Song2024frontiers}%
  \BibitemOpen
  \bibfield  {author} {\bibinfo {author} {\bibfnamefont {Y.}~\bibnamefont {Song}}, \bibinfo {author} {\bibfnamefont {J.}~\bibnamefont {Li}}, \bibinfo {author} {\bibfnamefont {Y.}~\bibnamefont {Wu}}, \bibinfo {author} {\bibfnamefont {S.}~\bibnamefont {Qin}}, \bibinfo {author} {\bibfnamefont {Q.}~\bibnamefont {Wen}},\ and\ \bibinfo {author} {\bibfnamefont {F.}~\bibnamefont {Gao}},\ }\bibfield  {journal} {\bibinfo  {journal} {Frontiers in Physics}\ }\textbf {\bibinfo {volume} {Volume 12 - 2024}},\ \href {https://doi.org/10.3389/fphy.2024.1362690} {10.3389/fphy.2024.1362690} (\bibinfo {year} {2024})\BibitemShut {NoStop}%
\bibitem [{\citenamefont {Goh}\ \emph {et~al.}(2023)\citenamefont {Goh}, \citenamefont {Larocca}, \citenamefont {Cincio}, \citenamefont {Cerezo},\ and\ \citenamefont {Sauvage}}]{goh2023liealgebraic}%
  \BibitemOpen
  \bibfield  {author} {\bibinfo {author} {\bibfnamefont {M.~L.}\ \bibnamefont {Goh}}, \bibinfo {author} {\bibfnamefont {M.}~\bibnamefont {Larocca}}, \bibinfo {author} {\bibfnamefont {L.}~\bibnamefont {Cincio}}, \bibinfo {author} {\bibfnamefont {M.}~\bibnamefont {Cerezo}},\ and\ \bibinfo {author} {\bibfnamefont {F.}~\bibnamefont {Sauvage}},\ }\href@noop {} {\bibinfo {title} {Lie-algebraic classical simulations for variational quantum computing}} (\bibinfo {year} {2023}),\ \Eprint {https://arxiv.org/abs/2308.01432} {arXiv:2308.01432 [quant-ph]} \BibitemShut {NoStop}%
\bibitem [{\citenamefont {Cerezo}\ \emph {et~al.}(2023)\citenamefont {Cerezo}, \citenamefont {Larocca}, \citenamefont {García-Martín}, \citenamefont {Diaz}, \citenamefont {Braccia}, \citenamefont {Fontana}, \citenamefont {Rudolph}, \citenamefont {Bermejo}, \citenamefont {Ijaz}, \citenamefont {Thanasilp}, \citenamefont {Anschuetz},\ and\ \citenamefont {Holmes}}]{Cerezo2023CSIM}%
  \BibitemOpen
  \bibfield  {author} {\bibinfo {author} {\bibfnamefont {M.}~\bibnamefont {Cerezo}}, \bibinfo {author} {\bibfnamefont {M.}~\bibnamefont {Larocca}}, \bibinfo {author} {\bibfnamefont {D.}~\bibnamefont {García-Martín}}, \bibinfo {author} {\bibfnamefont {N.~L.}\ \bibnamefont {Diaz}}, \bibinfo {author} {\bibfnamefont {P.}~\bibnamefont {Braccia}}, \bibinfo {author} {\bibfnamefont {E.}~\bibnamefont {Fontana}}, \bibinfo {author} {\bibfnamefont {M.~S.}\ \bibnamefont {Rudolph}}, \bibinfo {author} {\bibfnamefont {P.}~\bibnamefont {Bermejo}}, \bibinfo {author} {\bibfnamefont {A.}~\bibnamefont {Ijaz}}, \bibinfo {author} {\bibfnamefont {S.}~\bibnamefont {Thanasilp}}, \bibinfo {author} {\bibfnamefont {E.~R.}\ \bibnamefont {Anschuetz}},\ and\ \bibinfo {author} {\bibfnamefont {Z.}~\bibnamefont {Holmes}},\ }\href@noop {} {\bibinfo {title} {Does provable absence of barren plateaus imply classical simulability? or, why we need to rethink variational quantum computing}} (\bibinfo {year} {2023}),\ \Eprint
  {https://arxiv.org/abs/2312.09121} {arXiv:2312.09121 [quant-ph]} \BibitemShut {NoStop}%
\bibitem [{\citenamefont {Bermejo}\ \emph {et~al.}(2024)\citenamefont {Bermejo}, \citenamefont {Braccia}, \citenamefont {Rudolph}, \citenamefont {Holmes}, \citenamefont {Cincio},\ and\ \citenamefont {Cerezo}}]{bermejo2024qcnn}%
  \BibitemOpen
  \bibfield  {author} {\bibinfo {author} {\bibfnamefont {P.}~\bibnamefont {Bermejo}}, \bibinfo {author} {\bibfnamefont {P.}~\bibnamefont {Braccia}}, \bibinfo {author} {\bibfnamefont {M.~S.}\ \bibnamefont {Rudolph}}, \bibinfo {author} {\bibfnamefont {Z.}~\bibnamefont {Holmes}}, \bibinfo {author} {\bibfnamefont {L.}~\bibnamefont {Cincio}},\ and\ \bibinfo {author} {\bibfnamefont {M.}~\bibnamefont {Cerezo}},\ }\href {https://arxiv.org/abs/2408.12739} {\bibinfo {title} {Quantum convolutional neural networks are (effectively) classically simulable}} (\bibinfo {year} {2024}),\ \Eprint {https://arxiv.org/abs/2408.12739} {arXiv:2408.12739 [quant-ph]} \BibitemShut {NoStop}%
\bibitem [{\citenamefont {O'Shea}\ and\ \citenamefont {Nash}(2015)}]{oshea2015cnns}%
  \BibitemOpen
  \bibfield  {author} {\bibinfo {author} {\bibfnamefont {K.}~\bibnamefont {O'Shea}}\ and\ \bibinfo {author} {\bibfnamefont {R.}~\bibnamefont {Nash}},\ }\href {https://arxiv.org/abs/1511.08458} {\bibinfo {title} {An introduction to convolutional neural networks}} (\bibinfo {year} {2015}),\ \Eprint {https://arxiv.org/abs/1511.08458} {arXiv:1511.08458 [cs.NE]} \BibitemShut {NoStop}%
\bibitem [{\citenamefont {Devenport}\ \emph {et~al.}(1996)\citenamefont {Devenport}, \citenamefont {Rife}, \citenamefont {Liapis},\ and\ \citenamefont {Follin}}]{lamb_oseen_vortex}%
  \BibitemOpen
  \bibfield  {author} {\bibinfo {author} {\bibfnamefont {W.~J.}\ \bibnamefont {Devenport}}, \bibinfo {author} {\bibfnamefont {M.~C.}\ \bibnamefont {Rife}}, \bibinfo {author} {\bibfnamefont {S.~I.}\ \bibnamefont {Liapis}},\ and\ \bibinfo {author} {\bibfnamefont {G.~J.}\ \bibnamefont {Follin}},\ }\href {https://doi.org/10.1017/S0022112096001929} {\bibfield  {journal} {\bibinfo  {journal} {Journal of Fluid Mechanics}\ }\textbf {\bibinfo {volume} {312}},\ \bibinfo {pages} {67–106} (\bibinfo {year} {1996})}\BibitemShut {NoStop}%
\bibitem [{\citenamefont {Ramakrishnan}(1971)}]{ramakrishnan1971generalized}%
  \BibitemOpen
  \bibfield  {author} {\bibinfo {author} {\bibfnamefont {A.}~\bibnamefont {Ramakrishnan}},\ }in\ \href@noop {} {\emph {\bibinfo {booktitle} {Proceedings of the Conference on Clifford Algebra, its Generalization and Applications}}}\ (\bibinfo  {publisher} {Matscience},\ \bibinfo {address} {Madras},\ \bibinfo {year} {1971})\ pp.\ \bibinfo {pages} {87--96}\BibitemShut {NoStop}%
\bibitem [{\citenamefont {S{\"{u}}nderhauf}\ \emph {et~al.}(2024)\citenamefont {S{\"{u}}nderhauf}, \citenamefont {Campbell},\ and\ \citenamefont {Camps}}]{Sunderhauf2024blockencoding}%
  \BibitemOpen
  \bibfield  {author} {\bibinfo {author} {\bibfnamefont {C.}~\bibnamefont {S{\"{u}}nderhauf}}, \bibinfo {author} {\bibfnamefont {E.}~\bibnamefont {Campbell}},\ and\ \bibinfo {author} {\bibfnamefont {J.}~\bibnamefont {Camps}},\ }\href {https://doi.org/10.22331/q-2024-01-11-1226} {\bibfield  {journal} {\bibinfo  {journal} {{Quantum}}\ }\textbf {\bibinfo {volume} {8}},\ \bibinfo {pages} {1226} (\bibinfo {year} {2024})}\BibitemShut {NoStop}%
\bibitem [{\citenamefont {Camps}\ \emph {et~al.}(2023)\citenamefont {Camps}, \citenamefont {Lin}, \citenamefont {Beeumen},\ and\ \citenamefont {Yang}}]{camps2023explicitquantumcircuitsblock}%
  \BibitemOpen
  \bibfield  {author} {\bibinfo {author} {\bibfnamefont {D.}~\bibnamefont {Camps}}, \bibinfo {author} {\bibfnamefont {L.}~\bibnamefont {Lin}}, \bibinfo {author} {\bibfnamefont {R.~V.}\ \bibnamefont {Beeumen}},\ and\ \bibinfo {author} {\bibfnamefont {C.}~\bibnamefont {Yang}},\ }\href {https://arxiv.org/abs/2203.10236} {\bibinfo {title} {Explicit quantum circuits for block encodings of certain sparse matrices}} (\bibinfo {year} {2023}),\ \Eprint {https://arxiv.org/abs/2203.10236} {arXiv:2203.10236 [quant-ph]} \BibitemShut {NoStop}%
\bibitem [{\citenamefont {Budinski}\ \emph {et~al.}(2023)\citenamefont {Budinski}, \citenamefont {Niemimäki}, \citenamefont {Zamora-Zamora},\ and\ \citenamefont {Lahtinen}}]{Budinski_2023}%
  \BibitemOpen
  \bibfield  {author} {\bibinfo {author} {\bibfnamefont {L.}~\bibnamefont {Budinski}}, \bibinfo {author} {\bibfnamefont {O.}~\bibnamefont {Niemimäki}}, \bibinfo {author} {\bibfnamefont {R.}~\bibnamefont {Zamora-Zamora}},\ and\ \bibinfo {author} {\bibfnamefont {V.}~\bibnamefont {Lahtinen}},\ }\href {https://doi.org/10.1088/2058-9565/acfab7} {\bibfield  {journal} {\bibinfo  {journal} {Quantum Science and Technology}\ }\textbf {\bibinfo {volume} {8}},\ \bibinfo {pages} {045031} (\bibinfo {year} {2023})}\BibitemShut {NoStop}%
\bibitem [{\citenamefont {Shende}\ \emph {et~al.}(2006)\citenamefont {Shende}, \citenamefont {Prasad}, \citenamefont {Markov},\ and\ \citenamefont {Hayes}}]{Shende2006}%
  \BibitemOpen
  \bibfield  {author} {\bibinfo {author} {\bibfnamefont {V.~V.}\ \bibnamefont {Shende}}, \bibinfo {author} {\bibfnamefont {A.~K.}\ \bibnamefont {Prasad}}, \bibinfo {author} {\bibfnamefont {I.~L.}\ \bibnamefont {Markov}},\ and\ \bibinfo {author} {\bibfnamefont {J.~P.}\ \bibnamefont {Hayes}},\ }\href {https://doi.org/10.1109/TCAD.2003.811448} {\bibfield  {journal} {\bibinfo  {journal} {Trans. Comp.-Aided Des. Integ. Cir. Sys.}\ }\textbf {\bibinfo {volume} {22}},\ \bibinfo {pages} {710–722} (\bibinfo {year} {2006})}\BibitemShut {NoStop}%
\bibitem [{\citenamefont {Herbert}\ \emph {et~al.}(2024)\citenamefont {Herbert}, \citenamefont {Sorci},\ and\ \citenamefont {Tang}}]{Herbert2024}%
  \BibitemOpen
  \bibfield  {author} {\bibinfo {author} {\bibfnamefont {S.}~\bibnamefont {Herbert}}, \bibinfo {author} {\bibfnamefont {J.}~\bibnamefont {Sorci}},\ and\ \bibinfo {author} {\bibfnamefont {Y.}~\bibnamefont {Tang}},\ }\href {https://doi.org/10.1103/PhysRevA.110.012437} {\bibfield  {journal} {\bibinfo  {journal} {Phys. Rev. A}\ }\textbf {\bibinfo {volume} {110}},\ \bibinfo {pages} {012437} (\bibinfo {year} {2024})}\BibitemShut {NoStop}%
\bibitem [{\citenamefont {Wills}\ and\ \citenamefont {Strelchuk}(2024)}]{wills2024schur}%
  \BibitemOpen
  \bibfield  {author} {\bibinfo {author} {\bibfnamefont {A.}~\bibnamefont {Wills}}\ and\ \bibinfo {author} {\bibfnamefont {S.}~\bibnamefont {Strelchuk}},\ }\href {https://arxiv.org/abs/2305.04069} {\bibinfo {title} {Generalised coupling and an elementary algorithm for the quantum schur transform}} (\bibinfo {year} {2024}),\ \Eprint {https://arxiv.org/abs/2305.04069} {arXiv:2305.04069 [quant-ph]} \BibitemShut {NoStop}%
\bibitem [{\citenamefont {Nielsen}\ and\ \citenamefont {Chuang}(2010)}]{nielsen2010quantum}%
  \BibitemOpen
  \bibfield  {author} {\bibinfo {author} {\bibfnamefont {M.~A.}\ \bibnamefont {Nielsen}}\ and\ \bibinfo {author} {\bibfnamefont {I.~L.}\ \bibnamefont {Chuang}},\ }\href@noop {} {\emph {\bibinfo {title} {Quantum computation and quantum information}}}\ (\bibinfo  {publisher} {Cambridge university press},\ \bibinfo {year} {2010})\BibitemShut {NoStop}%
\bibitem [{\citenamefont {Bäumer}\ \emph {et~al.}(2025)\citenamefont {Bäumer}, \citenamefont {Sutter},\ and\ \citenamefont {Woerner}}]{baumer2025aqft}%
  \BibitemOpen
  \bibfield  {author} {\bibinfo {author} {\bibfnamefont {E.}~\bibnamefont {Bäumer}}, \bibinfo {author} {\bibfnamefont {D.}~\bibnamefont {Sutter}},\ and\ \bibinfo {author} {\bibfnamefont {S.}~\bibnamefont {Woerner}},\ }\href {https://arxiv.org/abs/2504.20832} {\bibinfo {title} {Approximate quantum fourier transform in logarithmic depth on a line}} (\bibinfo {year} {2025}),\ \Eprint {https://arxiv.org/abs/2504.20832} {arXiv:2504.20832 [quant-ph]} \BibitemShut {NoStop}%
\bibitem [{\citenamefont {Oszmaniec}\ \emph {et~al.}(2019)\citenamefont {Oszmaniec}, \citenamefont {Maciejewski},\ and\ \citenamefont {Puchała}}]{measurement_with_postselection}%
  \BibitemOpen
  \bibfield  {author} {\bibinfo {author} {\bibfnamefont {M.}~\bibnamefont {Oszmaniec}}, \bibinfo {author} {\bibfnamefont {F.~B.}\ \bibnamefont {Maciejewski}},\ and\ \bibinfo {author} {\bibfnamefont {Z.}~\bibnamefont {Puchała}},\ }\bibfield  {journal} {\bibinfo  {journal} {Physical Review A}\ }\textbf {\bibinfo {volume} {100}},\ \href {https://doi.org/10.1103/physreva.100.012351} {10.1103/physreva.100.012351} (\bibinfo {year} {2019})\BibitemShut {NoStop}%
\bibitem [{\citenamefont {Umeano}\ \emph {et~al.}(2024{\natexlab{d}})\citenamefont {Umeano}, \citenamefont {Scali},\ and\ \citenamefont {Kyriienko}}]{umeano2024community}%
  \BibitemOpen
  \bibfield  {author} {\bibinfo {author} {\bibfnamefont {C.}~\bibnamefont {Umeano}}, \bibinfo {author} {\bibfnamefont {S.}~\bibnamefont {Scali}},\ and\ \bibinfo {author} {\bibfnamefont {O.}~\bibnamefont {Kyriienko}},\ }\href {https://arxiv.org/abs/2412.13160} {\bibinfo {title} {Quantum community detection via deterministic elimination}} (\bibinfo {year} {2024}{\natexlab{d}}),\ \Eprint {https://arxiv.org/abs/2412.13160} {arXiv:2412.13160 [quant-ph]} \BibitemShut {NoStop}%
\bibitem [{\citenamefont {Lguensat}\ \emph {et~al.}(2018)\citenamefont {Lguensat}, \citenamefont {Sun}, \citenamefont {Fablet}, \citenamefont {Tandeo}, \citenamefont {Mason},\ and\ \citenamefont {Chen}}]{eddynet}%
  \BibitemOpen
  \bibfield  {author} {\bibinfo {author} {\bibfnamefont {R.}~\bibnamefont {Lguensat}}, \bibinfo {author} {\bibfnamefont {M.}~\bibnamefont {Sun}}, \bibinfo {author} {\bibfnamefont {R.}~\bibnamefont {Fablet}}, \bibinfo {author} {\bibfnamefont {P.}~\bibnamefont {Tandeo}}, \bibinfo {author} {\bibfnamefont {E.}~\bibnamefont {Mason}},\ and\ \bibinfo {author} {\bibfnamefont {G.}~\bibnamefont {Chen}},\ }in\ \href {https://doi.org/10.1109/IGARSS.2018.8518411} {\emph {\bibinfo {booktitle} {IGARSS 2018 - 2018 IEEE International Geoscience and Remote Sensing Symposium}}}\ (\bibinfo {year} {2018})\ pp.\ \bibinfo {pages} {1764--1767}\BibitemShut {NoStop}%
\bibitem [{\citenamefont {Gourianov}\ \emph {et~al.}(2022)\citenamefont {Gourianov}, \citenamefont {Lubasch}, \citenamefont {Dolgov}, \citenamefont {van~den Berg}, \citenamefont {Babaee}, \citenamefont {Givi}, \citenamefont {Kiffner},\ and\ \citenamefont {Jaksch}}]{Gourianov2022}%
  \BibitemOpen
  \bibfield  {author} {\bibinfo {author} {\bibfnamefont {N.}~\bibnamefont {Gourianov}}, \bibinfo {author} {\bibfnamefont {M.}~\bibnamefont {Lubasch}}, \bibinfo {author} {\bibfnamefont {S.}~\bibnamefont {Dolgov}}, \bibinfo {author} {\bibfnamefont {Q.~Y.}\ \bibnamefont {van~den Berg}}, \bibinfo {author} {\bibfnamefont {H.}~\bibnamefont {Babaee}}, \bibinfo {author} {\bibfnamefont {P.}~\bibnamefont {Givi}}, \bibinfo {author} {\bibfnamefont {M.}~\bibnamefont {Kiffner}},\ and\ \bibinfo {author} {\bibfnamefont {D.}~\bibnamefont {Jaksch}},\ }\href {https://doi.org/10.1038/s43588-021-00181-1} {\bibfield  {journal} {\bibinfo  {journal} {Nature Computational Science}\ }\textbf {\bibinfo {volume} {2}},\ \bibinfo {pages} {30} (\bibinfo {year} {2022})}\BibitemShut {NoStop}%
\bibitem [{\citenamefont {Lloyd}\ \emph {et~al.}(2016)\citenamefont {Lloyd}, \citenamefont {Garnerone},\ and\ \citenamefont {Zanardi}}]{Lloyd2016}%
  \BibitemOpen
  \bibfield  {author} {\bibinfo {author} {\bibfnamefont {S.}~\bibnamefont {Lloyd}}, \bibinfo {author} {\bibfnamefont {S.}~\bibnamefont {Garnerone}},\ and\ \bibinfo {author} {\bibfnamefont {P.}~\bibnamefont {Zanardi}},\ }\bibfield  {journal} {\bibinfo  {journal} {Nature Communications}\ }\textbf {\bibinfo {volume} {7}},\ \href {https://doi.org/10.1038/ncomms10138} {10.1038/ncomms10138} (\bibinfo {year} {2016})\BibitemShut {NoStop}%
\bibitem [{\citenamefont {Gyurik}\ \emph {et~al.}(2022)\citenamefont {Gyurik}, \citenamefont {Cade},\ and\ \citenamefont {Dunjko}}]{Gyurik2022}%
  \BibitemOpen
  \bibfield  {author} {\bibinfo {author} {\bibfnamefont {C.}~\bibnamefont {Gyurik}}, \bibinfo {author} {\bibfnamefont {C.}~\bibnamefont {Cade}},\ and\ \bibinfo {author} {\bibfnamefont {V.}~\bibnamefont {Dunjko}},\ }\href {https://doi.org/10.22331/q-2022-11-10-855} {\bibfield  {journal} {\bibinfo  {journal} {Quantum}\ }\textbf {\bibinfo {volume} {6}},\ \bibinfo {pages} {855} (\bibinfo {year} {2022})}\BibitemShut {NoStop}%
\bibitem [{\citenamefont {Berry}\ \emph {et~al.}(2024)\citenamefont {Berry}, \citenamefont {Su}, \citenamefont {Gyurik}, \citenamefont {King}, \citenamefont {Basso}, \citenamefont {Barba}, \citenamefont {Rajput}, \citenamefont {Wiebe}, \citenamefont {Dunjko},\ and\ \citenamefont {Babbush}}]{Berry2024}%
  \BibitemOpen
  \bibfield  {author} {\bibinfo {author} {\bibfnamefont {D.~W.}\ \bibnamefont {Berry}}, \bibinfo {author} {\bibfnamefont {Y.}~\bibnamefont {Su}}, \bibinfo {author} {\bibfnamefont {C.}~\bibnamefont {Gyurik}}, \bibinfo {author} {\bibfnamefont {R.}~\bibnamefont {King}}, \bibinfo {author} {\bibfnamefont {J.}~\bibnamefont {Basso}}, \bibinfo {author} {\bibfnamefont {A.~D.~T.}\ \bibnamefont {Barba}}, \bibinfo {author} {\bibfnamefont {A.}~\bibnamefont {Rajput}}, \bibinfo {author} {\bibfnamefont {N.}~\bibnamefont {Wiebe}}, \bibinfo {author} {\bibfnamefont {V.}~\bibnamefont {Dunjko}},\ and\ \bibinfo {author} {\bibfnamefont {R.}~\bibnamefont {Babbush}},\ }\bibfield  {journal} {\bibinfo  {journal} {PRX Quantum}\ }\textbf {\bibinfo {volume} {5}},\ \href {https://doi.org/10.1103/prxquantum.5.010319} {10.1103/prxquantum.5.010319} (\bibinfo {year} {2024})\BibitemShut {NoStop}%
\bibitem [{\citenamefont {Scali}\ \emph {et~al.}(2024{\natexlab{a}})\citenamefont {Scali}, \citenamefont {Umeano},\ and\ \citenamefont {Kyriienko}}]{Scali2024dos}%
  \BibitemOpen
  \bibfield  {author} {\bibinfo {author} {\bibfnamefont {S.}~\bibnamefont {Scali}}, \bibinfo {author} {\bibfnamefont {C.}~\bibnamefont {Umeano}},\ and\ \bibinfo {author} {\bibfnamefont {O.}~\bibnamefont {Kyriienko}},\ }\bibfield  {journal} {\bibinfo  {journal} {Physical Review A}\ }\textbf {\bibinfo {volume} {110}},\ \href {https://doi.org/10.1103/physreva.110.042616} {10.1103/physreva.110.042616} (\bibinfo {year} {2024}{\natexlab{a}})\BibitemShut {NoStop}%
\bibitem [{\citenamefont {Scali}\ \emph {et~al.}(2024{\natexlab{b}})\citenamefont {Scali}, \citenamefont {Umeano},\ and\ \citenamefont {Kyriienko}}]{Scali2024thermal}%
  \BibitemOpen
  \bibfield  {author} {\bibinfo {author} {\bibfnamefont {S.}~\bibnamefont {Scali}}, \bibinfo {author} {\bibfnamefont {C.}~\bibnamefont {Umeano}},\ and\ \bibinfo {author} {\bibfnamefont {O.}~\bibnamefont {Kyriienko}},\ }\href {https://doi.org/10.1063/5.0209201} {\bibfield  {journal} {\bibinfo  {journal} {APL Quantum}\ }\textbf {\bibinfo {volume} {1}},\ \bibinfo {pages} {036106} (\bibinfo {year} {2024}{\natexlab{b}})}\BibitemShut {NoStop}%
\end{thebibliography}
%

\end{document}